\documentclass[a4paper,twoside,twocolumn,english,aps,pre,showpacs]{revtex4}
\usepackage{ae}
\usepackage{aecompl}
\usepackage[T1]{fontenc}
\usepackage[latin1]{inputenc}
\usepackage{amsmath}
\usepackage{graphicx}
\usepackage{amssymb}

\makeatletter

\newcommand{\noun}[1]{\textsc{#1}}
\providecommand{\tabularnewline}{\\}

\newcommand{\erfc}{\mathop{\operator@font erfc}\nolimits}
\newcommand{\sgn}{\mathop{\operator@font sgn}\nolimits}
\def\vec{\mathbf}
\newcommand{\C}[1]{{}} 
\newcommand{\bS}{\begin{subequations}}
\newcommand{\eS}{\end{subequations}}
\newcommand{\eg}{e.\,g.\ }
\newcommand{\ie}{i.\,e.\ }

\usepackage{babel}
\makeatother
\begin{document}

\newcommand{\para}{||}

\renewcommand{\para}{\|}

\title{Universal finite-size scaling analysis of Ising models with long-range
interactions at the upper critical dimensionality: Isotropic case}

\author{Daniel Grüneberg}

\email{daniel@thp.Uni-Duisburg.de}

\author{Alfred Hucht}

\email{fred@thp.Uni-Duisburg.de}

\affiliation{Fakultät für Naturwissenschaften, Theoretische Physik, Universität
Duisburg-Essen, D-47048 Duisburg, Germany}

\begin{abstract}
We investigate a two-dimensional Ising model with long-range interactions
that emerge from a generalization of the magnetic dipolar interaction
in spin systems with in-plane spin orientation. This interaction is,
in general, anisotropic whereby in the present work we focus on the
isotropic case for which the model is found to be at its upper critical
dimensionality. To investigate the critical behavior the temperature
and field dependence of several quantities are studied by means of
Monte Carlo simulations. On the basis of the Privman-Fisher hypothesis
and results of the renormalization group the numerical data are analyzed
in the framework of a finite-size scaling analysis and compared to
finite-size scaling functions derived from a Ginzburg-Landau-Wilson
model in zero mode (mean-field) approximation. The obtained excellent
agreement suggests that at least in the present case the concept of
universal finite-size scaling functions can be extended to the upper
critical dimensionality. 
\end{abstract}


\maketitle
\markboth{\rm{\it Physical Review} E {\bf 69}, 036104 (2004)}{\rm{\it Physical Review} E {\bf 69}, 036104 (2004)}

\section{Introduction\label{sec:Introduction}}

In the last decade, spin models with long-range interactions were
the subject of several extensive Monte Carlo studies. Utilizing an
efficient cluster algorithm \cite{LuijtenBloete95} these studies
were addressed to the verification of some unproved predictions on
the critical behavior of spin models with algebraically decaying long-range
interactions \cite{LuijtenBloete97}. Furthermore, the crossover from
Ising-like to classical critical behavior was investigated \cite{LuiBloeBin96,LuiBloeBin97}
and first numerical results on the critical behavior of the dipolar
in-plane Ising (DIPI) model were obtained \cite{Hucht02a}. This two-dimensional
model displays a strongly anisotropic phase transition, \ie, the
correlation lengths in direction parallel and perpendicular to spin
orientation diverge in the infinite system (let $t>0$) as \cite{Hucht02a}\begin{equation}
\xi_{\parallel}^{(\infty)}(t)\sim\hat{\xi}_{\parallel}t^{-\nu_{\parallel}},\qquad\xi_{\perp}^{(\infty)}(t)\sim\hat{\xi}_{\perp}t^{-\nu_{\perp}}\end{equation}
at the critical point, where both $\hat{\xi}_{\parallel}\neq\hat{\xi}_{\perp}$
and $\nu_{\parallel}\neq\nu_{\perp}$, and $t\equiv(T-T_{\mathrm{c}})/T_{\mathrm{c}}$
denotes the reduced temperature. Except the anisotropy exponent $\theta=\nu_{\parallel}/\nu_{\perp}$
neither any numerical estimates of the critical exponents exist for
the DIPI model, nor is it clear whether the model exhibits Lifshitz
type critical behavior \cite{Hornreich75} as it is observed, \eg,
in the anisotropic next nearest neighbor Ising (ANNNI) model \cite{Selke92,DiehlShpot00,PleimlingHenkel01}. 

\begin{figure}[t]
\begin{center}\includegraphics[%
  clip,
  width=6.5cm,
  keepaspectratio]{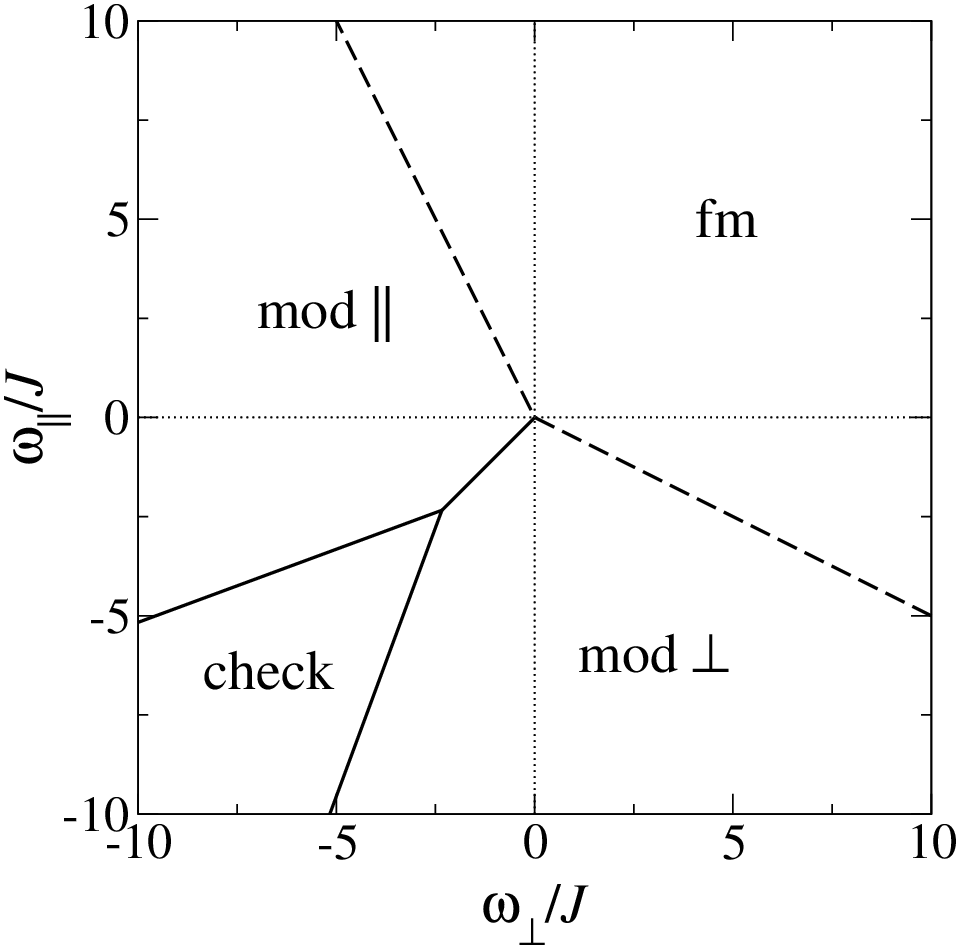}\end{center}

\caption{The model (Eqs.~(\ref{eqs:Model})) exhibits a ferromagnetic ground
state in the region denoted by {}``fm''. The dashed lines $\omega_{\para}=-2\omega_{\bot}>0$
and $-2\omega_{\para}=\omega_{\bot}>0$, respectively, and the origin
$\omega_{\parallel}=\omega_{\perp}=0$ also represent ferromagnetic
ground states. All other regions correspond to modulated spin configurations
(see text). \label{cap:Phasediagram}}
\end{figure}
To address such questions in a broader context we present a two-dimensional
long-range spin model that arises from generalizing the DIPI model
by introducing an additional parameter. Assuming $L_{\para}\times L_{\bot}$
geometry and periodic boundary conditions this model is described
by the Hamiltonian \bS\label{eqs:Model}\begin{equation}
\mathcal{H}=-\frac{1}{2}\sum_{i\neq j}s_{i}J(\vec{r}_{ij})s_{j}-B\sum_{i}s_{i}\label{Hamiltonian}\end{equation}
with magnetic Ising spin variables $s_{i}=\pm1$, the spin-spin distance
vector $\vec{r}_{ij}$, and an external field $B$. The pair coupling
$J(\vec{r})$ is given by\begin{equation}
J(\vec{r})=J\delta_{|\vec{r}|,1}+\frac{\omega_{\para}r_{\para}^{2}+\omega_{\bot}r_{\bot}^{2}}{|\vec{r}|^{5}}\label{Interaction}\end{equation}
\eS and consists of both a ferromagnetic short-range nearest neighbor
exchange coupling with the coupling constant $J\geq0$ and a long-range
contribution, where $r_{\para}$ and $r_{\bot}$ are the components
of the vector $\vec{r}$ parallel and perpendicular to spin orientation.
Using this general form of the pair coupling $J(\vec{r})$, several
well known spin systems can be mapped onto this model by making an
explicit choice of the parameters $\omega_{\para}$, $\omega_{\bot}$,
and $J$. With $\omega_{\para}=-2\omega_{\bot}>0$, and for symmetry
reasons $-2\omega_{\para}=\omega_{\bot}>0$, and $J>0$ the DIPI model
is recovered, and if $\omega_{\para}=\omega_{\bot}>0$ and $J=0$
Eq.~(\ref{Hamiltonian}) corresponds to an Ising model with an isotropic
ferromagnetic long-range interaction algebraically decaying as $J(\vec{r})\propto|\vec{r}|^{-3}$.
Another special case is the dipolar Ising model with perpendicular
spin orientation \cite{Taylor93,MacIsaac95} that can be obtained
for $\omega_{\para}=\omega_{\bot}<0$ and $J>0$. 

Figure~\ref{cap:Phasediagram} shows the ground state phase diagram
of the model whereby we took into account four different ground state
spin configurations: the \textit{ferromagnetic state} (fm) where all
spins point to the same direction, the \textit{totally antiferromagnetic
state} that is referred to as checkerboard state (check), and \textit{commensurate
stripe domain states} with a domain wall orientation parallel (mod
$\parallel$) and perpendicular (mod $\perp$) to spin orientation
and the periods $N_{\parallel,\perp}$. Dependent on the values of
the quotients $\omega_{\parallel}/J$ and $\omega_{\perp}/J$ all
considered spin configurations were found as stable ground states.
Due to symmetry reasons the arrangement of the corresponding phases
in Fig.~\ref{cap:Phasediagram} is symmetric with respect to the
line $\omega_{\parallel}=\omega_{\perp}$. The periods $N_{\parallel,\perp}$
of the stripe domain states diverge when approaching the dashed lines
in Fig.~\ref{cap:Phasediagram}. 

The region in parameter space where the model displays a ferromagnetic
ground state is of particular interest to us. In that region the observed
phase transitions are isotropic (anisotropic) when $\omega_{\parallel}=\omega_{\perp}$
($\omega_{\parallel}\neq\omega_{\perp}$) whereby in the present work
we draw our attention to the isotropic ferromagnetic long-range case\begin{equation}
\omega_{\parallel}=\omega_{\perp}>0,\qquad J=0\label{eq:isolrfmcase}\end{equation}
before we turn to the case of an anisotropic pair coupling \cite{HuchtGrueneberg04}.
It is known that the upper critical dimension of long-range spin models
with ferromagnetic interactions decaying as $J(\vec{r})\propto|\vec{r}|^{-(d+\sigma)}$
is given by $d_{\mathrm{u}}=2\sigma$ \cite{FisherMaNickel72}. Comparison
with the pair coupling $J(\vec{r})$ defined in Eq.~(\ref{Interaction})
yields in the isotropic case, Eq.~(\ref{eq:isolrfmcase}), $\sigma=1$
for a two-dimensional system and consequently $d=d_{\mathrm{u}}=2$. 

So in this paper we investigate the critical behavior of the model
at its borderline dimensionality $d_{\mathrm{u}}$ by means of Monte
Carlo (MC) simulations and finite-size scaling methods. For that purpose
in the following Section~\ref{sec:Finite-Size-Scaling-Analysis}
the finite-size scaling form of the free energy density is discussed
and the finite-size scaling relations of the considered quantities
are derived as they are used for the finite-size scaling analysis.
These relations define finite-size scaling functions for which in
Section~\ref{sec:Mean-Field-Theory} we evaluate analytical expressions
in the framework of the so-called zero mode theory that is based on
the Ginzburg-Landau-Wilson (GLW) model. In the last Section~\ref{sec:Monte-Carlo-Results}
the zero mode results are compared to numerical data within a finite-size
scaling analysis.

\section{Finite-Size Scaling Relations \label{sec:Finite-Size-Scaling-Analysis}}

To study the critical properties of the model in the isotropic long-range
case we have carried out a finite-size scaling analysis of MC data.
This analysis requires the finite-size scaling relations of the quantities
that were considered in the simulations. 

Via a renormalization group approach, Luijten and Blöte \cite{LuijtenBloete97}
derived the scaling form of the free energy density of $O(n)$ spin
models with ferromagnetic long-range interactions decaying as $J(\vec{r})\propto|\vec{r}|^{-(d+\sigma)}$.
At the upper critical dimension, that is given by $d_{\mathrm{u}}=2$
for $\sigma=1$ (see Sec.~\ref{sec:Introduction}), the singular
part of the reduced free energy density was found to scale as $(n=1)${\small \begin{eqnarray}
f_{s}(t,h;L) & \sim & L^{-2}\widetilde{f}\!\left(u^{-\frac{1}{3}}L\ln^{\frac{1}{6}}({\textstyle \frac{L}{L'_{0}}})\left[t-v'L^{-1}\ln^{-\frac{2}{3}}({\textstyle \frac{L}{L'_{0}}})\right],\right.\nonumber \\
 &  & \left.L^{\frac{3}{2}}\ln^{\frac{1}{4}}({\textstyle \frac{L}{L'_{0}}})h\vphantom{\left[t-v'L^{-1}\ln^{-\frac{2}{3}}({\textstyle \frac{L}{L'_{0}}})\right]}\right)\label{FSSRFED}\end{eqnarray}
}with the reduced temperature $t$, the reduced external field $h\equiv\beta B$
where $\beta\equiv1/(k_{\mathrm{B}}T)$ denotes the inverse temperature
(we set $k_{\mathrm{B}}{=}1$ throughout this paper), and the so-called
dangerous irrelevant variable $u$ \cite{Fisher82,PrivmanFisher83}.
Note that we rewrote the formula given in Ref.~\cite{LuijtenBloete97}
in terms of the parameters $v^{\prime}$ and $L_{0}^{\prime}$, where
$L_{0}^{\prime}$ can be regarded as a reference length that fixes
the length scale in the logarithms (see also Ref.~\cite{LuebeckHeger2003a}),
and we absorbed constant factors into $\widetilde{f}$. The symbol
$\sim$ means {}``asymptotically equal'' and, unless stated otherwise,
refers to the limit $(t,h,L)\rightarrow(0,0,\infty)$ with $tL\ln^{\frac{1}{6}}({\textstyle L})$
and $hL^{\frac{3}{2}}\ln^{\frac{1}{4}}(L)$ fixed (cf.~Eq.~(\ref{eq:expansions})). 

Proceeding from Eq.~(\ref{FSSRFED}) we adapt the Privman-Fisher
hypothesis \cite{PrivmanFisher84} and propose the finite-size scaling
form of the singular part of the reduced free energy density\begin{equation}
f_{s}(t,h;L)\sim L^{-2}Y(x_{\mathrm{rg}},y_{\mathrm{rg}})\label{eq:fs(t,h,L)}\end{equation}
with the universal finite-size scaling (UFSS) function $Y(x,y)$.
The arguments of this function correspond to the temperature scaling
variable \bS  \label{eqs:xy_rg}\begin{equation}
x_{\mathrm{rg}}=C_{1}\hat{t}L\ln^{\frac{1}{6}}({\textstyle \frac{L}{L_{0}}})\label{eq:xrg}\end{equation}
 with the shifted reduced temperature\begin{equation}
\hat{t}=t-vL^{-1}\ln^{-\frac{2}{3}}({\textstyle \frac{L}{L_{0}}}),\label{eq:dfdffd}\end{equation}
and the field scaling variable\begin{equation}
y_{\mathrm{rg}}=C_{2}hL^{\frac{3}{2}}\ln^{\frac{1}{4}}({\textstyle \frac{L}{L_{0}}}),\end{equation}
\eS  whereby $C_{1}$ and $C_{2}$ are nonuniversal metric factors.
Let us note that we have replaced the constants $v'$ and $L'_{0}$
which arise from the renormalization group by the constants $v$ and
$L_{0}$ that will be used as fit parameters in the finite-size scaling
analysis. 

It is also important to point out that the terms that result from
$v$ and $L_{0}$ in the temperature scaling variable $x_{\mathrm{rg}}$
and the field scaling variable $y_{\mathrm{rg}}$ are merely corrections
since they do not contribute to the leading orders in the expansions
\bS  \label{eq:expansions}\begin{eqnarray}
\hat{t}L\ln^{\frac{1}{6}}({\textstyle \frac{L}{L_{0}}}) & \stackrel{L\rightarrow\infty}{=} & tL\ln^{\frac{1}{6}}({\textstyle L})+v\mathcal{O}(\ln^{-\frac{1}{2}}(L))\nonumber \\
 &  & +tL\ln(L_{0})\mathcal{O}(\ln^{-\frac{5}{6}}({\textstyle L}))\\
hL^{\frac{3}{2}}\ln^{\frac{1}{4}}({\textstyle \frac{L}{L_{0}}}) & \stackrel{L\rightarrow\infty}{=} & hL^{\frac{3}{2}}\ln^{\frac{1}{4}}(L)\nonumber \\
 &  & +hL^{\frac{3}{2}}\ln(L_{0})\mathcal{O}(\ln^{-\frac{3}{4}}({\textstyle L})).\end{eqnarray}
\eS However, due to the slow convergence of the logarithms appearing
at the upper critical dimension these corrections are substantial
for the quality of the data collapse in the finite-size scaling analysis,
as will be discussed in Sec.~\ref{sec:Monte-Carlo-Results} (see
also Sec.~IV(B) in \cite{LuijtenBloete97}, and \cite{LuebeckHeger2003a}). 

In the following we derive the finite-size scaling forms of the quantities
that were considered in the simulations. Let\begin{equation}
\bar{s}=\frac{1}{L^{d}}\sum_{j=1}^{L^{d}}s_{j}\label{eq:avgspin}\end{equation}
denote the average of the spin variables $s_{j}$, these quantities
are the magnetization $m(t,h;L)=\langle\bar{s}\rangle$ and the susceptibility
$\chi(t,h;L)=\beta L^{d}(\langle\bar{s}^{2}\rangle-\langle\bar{s}\rangle^{2})$,
whose finite-size scaling forms can be obtained by taking derivatives
of the singular part of the reduced free energy density, Eq.~(\ref{eq:fs(t,h,L)}),
according to\bS \label{mchifss}\begin{eqnarray}
m(t,h;L) & = & -\frac{\partial}{\partial h}f_{s}(t,h;L)\nonumber \\
 & \sim & C_{2}L^{-\frac{1}{2}}\ln^{\frac{1}{4}}({\textstyle \frac{L}{L_{0}}})Y_{m}(x_{\mathrm{rg}},y_{\mathrm{rg}}),\label{eq:m(t,h,L)}\\
\beta^{-1}\chi(t,h;L) & = & -\frac{\partial^{2}}{\partial h^{2}}f_{s}(t,h;L)\nonumber \\
 & \sim & C_{2}^{2}L\ln^{\frac{1}{2}}({\textstyle \frac{L}{L_{0}}})Y_{\chi}(x_{\mathrm{rg}},y_{\mathrm{rg}}).\label{eq:chi(t,h,L)}\end{eqnarray}
\eS  We also consider the dimensionless Binder cumulant \cite{Binder81}
$U(t;L)=1-\langle\bar{s}^{4}\rangle/(3\langle\bar{s}^{2}\rangle^{2})$
that can be evaluated from the susceptibility $\chi(t,h;L)$ and the
nonlinear susceptibility $\chi^{\left(\mathrm{nl}\right)}(t,h;L)$
using the identity \cite{BinNauPrivYoung85}\begin{equation}
U(t;L)=-\frac{\chi^{\left(\mathrm{nl}\right)}(t,0;L)}{3\beta L^{2}\chi^{2}(t,0;L)},\end{equation}
where $\chi^{\left(\mathrm{nl}\right)}(t,h;L)$ is given by\begin{eqnarray}
\beta^{-3}\chi^{\left(\mathrm{nl}\right)}(t,h;L) & = & -\frac{\partial^{4}}{\partial h^{4}}f_{s}(t,h;L)\nonumber \\
 & \sim & C_{2}^{4}L^{4}\ln({\textstyle \frac{L}{L_{0}}})Y_{\chi^{\left(\mathrm{nl}\right)}}(x_{\mathrm{rg}},y_{\mathrm{rg}}).\label{eq:chinl(t,h,L)}\end{eqnarray}
Hence, the Binder cumulant scales as\begin{eqnarray}
U(t;L) & \sim & Y_{U}(x_{\mathrm{rg}}),\label{eq:U(t,L)}\end{eqnarray}
where $Y_{U}(x)\equiv-Y_{\chi^{\left(\mathrm{nl}\right)}}(x,0)/(3Y_{\chi}^{2}(x,0))$.
Since the ensemble averages $\langle\bar{s}^{2m+1}\rangle$ with $m\in\mathbb{N}_{0}$
vanish in the absence of an external field, we, in addition, analyze
the magnetization $m_{\mathrm{abs}}(t,h;L)=\langle|\bar{s}|\rangle$
and the susceptibility $\chi_{\mathrm{abs}}(t,h;L)=\beta L^{d}(\langle\bar{s}^{2}\rangle-\langle|\bar{s}|\rangle^{2})$.
It is understood that these quantities also fulfill the finite-size
scaling forms of $m(t,h;L)$ and $\chi(t,h;L)$, Eqs.~(\ref{mchifss}),
respectively with the corresponding UFSS functions $Y_{m_{\mathrm{abs}}}(x,y)$
and $Y_{\chi_{\mathrm{abs}}}(x,y)$.

\section{Mean-Field Theory \label{sec:Mean-Field-Theory}}

An appropriate description of spin systems of dimensions $L\times L\times\cdots\times L=L^{d}$
with mean-field-like (classical) critical behavior can be achieved
by the mean-field theory employed by Brézin \noun{}and Zinn-Justin~\cite{BrezinZinnJustin85}
and Rudnick \noun{}\textit{et al}\noun{.~\cite{RudnickGuoJasnow85}.}
This theory, also known as zero mode approximation, yields in contrast
to conventional mean-field theories a rounded transition for finite
systems. In the thermodynamic limit $L\rightarrow\infty$ the usual
power laws with the expected mean-field values of the critical exponents
can be recovered. 

In the following, this theory is reviewed and used to evaluate analytical
expressions for the finite-size scaling functions defined in the preceding
section in order to compare them to the numerical data, as it is demonstrated
in Sec.~\ref{sec:Monte-Carlo-Results}. The basis of this evaluation
is the reduced GLW Hamiltonian in momentum space that corresponds
to the underlying spin system with a long-range interaction $J(\vec{r})\propto|\vec{r}|^{-(d+\sigma)}$.
It is given by (see, \eg, Ref.~\cite{Aharony76})\begin{eqnarray}
\bar{\mathcal{H}} & = & L^{d}\!\left(\frac{1}{2}\sum_{\vec{k}}(r+A_{\sigma}|\vec{k}|^{\sigma})\varphi_{\vec{k}}\varphi_{-\vec{k}}-h\varphi_{\vec{0}}\right.\nonumber \\
 &  & \left.+\frac{u}{4!}\sum_{\vec{k}_{1}}\sum_{\vec{k}_{2}}\sum_{\vec{k}_{3}}\varphi_{\vec{k}_{1}}\varphi_{\vec{k}_{2}}\varphi_{\vec{k}_{3}}\varphi_{-\vec{k}_{1}-\vec{k}_{2}-\vec{k}_{3}}\right)\label{eq:H(phik)}\end{eqnarray}
with the temperature-like parameter $r\propto T-T_{\mathrm{c}}^{\mathrm{mf}}$
that measures the deviation of the temperature from the mean-field
critical temperature $T_{\mathrm{c}}^{\mathrm{mf}}$, the reduced
external field $h$, and the dangerous irrelevant variable $u>0.$
Each sum in Eq.~(\ref{eq:H(phik)}) runs for each component $k_{j}$
of $\vec{k}$ over integer multiples of $2\pi/L$ up to a momentum
space cutoff $k_{\Lambda}=\pi/a$ $(|k_{j}|\leq k_{\Lambda})$ with
the lattice constant~$a$. 

The essential step of the zero mode approximation is the neglection
of all modes except the zero mode $\varphi\equiv\varphi_{\vec{0}}$
in Eq.~(\ref{eq:H(phik)}). This leads to the reduced zero mode Hamiltonian
(see, \eg, Ref.~\cite{RudnickGuoJasnow85})\begin{equation}
\bar{\mathcal{H}}_{0}(\varphi)=L^{d}\!\left(\frac{r}{2}\varphi^{2}+\frac{u}{4!}\varphi^{4}-h\varphi\right).\label{eq:H0}\end{equation}
with the corresponding partition function\begin{equation}
\mathcal{Z}_{0}=\int_{-\infty}^{\infty}\mathrm{d}\varphi\, e^{-\bar{\mathcal{H}}_{0}(\varphi)}.\label{eq:ZMPF}\end{equation}
So the normalized probability distribution of the zero mode is given
by\begin{equation}
\mathcal{P}_{0}(\varphi)=\frac{1}{\mathcal{Z}_{0}}e^{-\bar{\mathcal{H}}_{0}(\varphi)}\label{zmaGibbsdistr}\end{equation}
and it can be used to evaluate averages of the form\begin{equation}
\langle g(\varphi)\rangle_{0}=\int_{-\infty}^{\infty}\mathrm{d}\varphi\, g(\varphi)\mathcal{P}_{0}(\varphi).\label{eq:avgzma}\end{equation}
A further central quantity is the reduced free energy density that
is given by $f_{0}=-L^{-d}\ln(\mathcal{Z}_{0})$ within the zero mode
approximation \cite{BinNauPrivYoung85,BrezinZinnJustin85}. Using
this expression and the zero mode partition function defined in Eq.~(\ref{eq:ZMPF}),
the rescaling (see, \eg, Ref.~\cite{BrezinZinnJustin85})\begin{equation}
\varphi\rightarrow(uL^{d})^{-\frac{1}{4}}\varphi\label{eq:rescalingII}\end{equation}
immediately yields the zero mode finite-size scaling form of $f_{0}$.
It reads\begin{equation}
f_{0}(r,h;L)=L^{-d}\tilde{f}_{0}(x_{\mathrm{mf}},y_{\mathrm{mf}})+c(L)\end{equation}
 with the mean-field temperature and field scaling variables \begin{equation}
x_{\mathrm{mf}}=ru^{-\frac{1}{2}}L^{\frac{d}{2}},\qquad y_{\mathrm{mf}}=hu^{-\frac{1}{4}}L^{\frac{3d}{4}},\label{eq:xy_mf}\end{equation}
 and an additive term $c(L)$ that is without significance in the
following since it is absent after taking derivatives of $f_{0}(r,h;L)$
with respect to $r$ or $h$. Instead we focus on the finite-size
scaling function\begin{equation}
\tilde{f}_{0}(x,y)=-\ln\left(\int_{-\infty}^{\infty}\mathrm{d}\varphi\, e^{-(\frac{x}{2}\varphi^{2}+\frac{1}{24}\varphi^{4}-y\varphi)}\right)\label{eq:FSSFSF}\end{equation}
from which, as seen in the preceding section, the finite-size scaling
functions of other quantities like the magnetization $m$ and the
susceptibility $\chi$ follow. 

The asymptotics of this function are given by \bS  \label{eq:asympt}\begin{eqnarray}
\tilde{f}_{0}(x,0) & \stackrel{x\rightarrow-\infty}{\sim} & -\frac{3}{2}x^{2}+\frac{1}{2}\ln\!\left(\frac{|x|}{4\pi}\right)\label{eq:asympt1}\\
\tilde{f}_{0}(x,0) & \stackrel{x\rightarrow+\infty}{\sim} & \frac{1}{2}\ln\!\left(\frac{x}{2\pi}\right)\\
\tilde{f}_{0}(0,y) & \stackrel{y\rightarrow\pm\infty}{\sim} & -\left(\frac{81}{32}\right)^{\frac{1}{3}}|y|^{\frac{4}{3}}.\label{eq:asympt2}\end{eqnarray}
\eS Following the convention suggested in Ref.~\cite{GlassPrivSchul87},
the normalized finite-size scaling function $Y^{\mathrm{mf}}(x,y)$
of the reduced free energy density should be defined such that their
asymptotics read\bS \begin{eqnarray}
Y^{\mathrm{mf}}(x,0) & \stackrel{x\rightarrow-\infty}{\sim} & -x^{2}\\
Y^{\mathrm{mf}}(0,y) & \stackrel{y\rightarrow\pm\infty}{\sim} & -|y|^{\frac{4}{3}}\end{eqnarray}
\eS instead of the leading orders in Eqs.~(\ref{eq:asympt}). This
requirement fixes some arbitrariness of the reduced free energy density
finite-size scaling function and can be fulfilled by a rescaling of
the parameters $r$ and $u$, which are the only phenomenological
quantities entering the reduced zero mode Hamiltonian, Eq.~(\ref{eq:H0}).
Replacing $x$ and $y$ in Eqs.~(\ref{eq:asympt}) explicitly with
the mean-field scaling variables $x_{\mathrm{mf}}$ and $y_{\mathrm{mf}}$
from Eqs. (\ref{eq:xy_mf}), the rescaling\begin{equation}
r\rightarrow\frac{3\sqrt{3}}{4}r,\qquad u\rightarrow\frac{81}{32}u\label{eq:rescaling}\end{equation}
 leads to a cancellation of the corresponding coefficients of $-x_{\mathrm{mf}}^{2}$
and $-|y_{\mathrm{mf}}|^{\frac{4}{3}}$ and one obtains the desired
asymptotics. It is important to note that we do not alter the definitions
of $x_{\mathrm{mf}}$ and $y_{\mathrm{mf}}$ due to this rescaling,
but the scaling function itself. Starting from the zero mode partition
function, Eq.~(\ref{eq:ZMPF}), with the rescaled parameters $r$
and $u$, we, after the procedure discussed above, finally end up
with\begin{equation}
Y^{\mathrm{mf}}(x,y)=-\ln(\Xi_{0}(x,y)+\Xi_{0}(x,-y))\end{equation}
where\begin{equation}
\Xi_{m}(x,y)=\int_{0}^{\infty}\mathrm{d}\varphi\,\varphi^{m}e^{-(\frac{3\sqrt{3}}{8}x\varphi^{2}+\frac{27}{256}\varphi^{4}-y\varphi)},\label{eq:Xi(x,y)}\end{equation}
instead of $\tilde{f}_{0}(x,y)$ given in Eq.~(\ref{eq:FSSFSF}). 

Having defined the reduced free energy density finite-size scaling
function, in the following the finite-size scaling forms and the corresponding
zero mode finite-size scaling functions of the quantities considered
in the preceding section will be derived. Since the quantities $m_{\mathrm{abs}}(t,h;L)$
and $\chi_{\mathrm{abs}}(t,h;L)$ cannot be evaluated by taking derivatives
of the reduced free energy density, we instead make use of the average
defined in Eq.~(\ref{eq:avgzma}). Setting here $g(\varphi)=\varphi^{m}$
and $g(\varphi)=|\varphi|^{m}$, respectively, the rescaling of the
parameters $r$ and $u$ {[}Eq.~(\ref{eq:rescaling}){]} and the
rescaling of the zero mode (Eq.~(\ref{eq:rescalingII})) immediately
yields the finite-size scaling forms of these averages. They read\bS \label{fssforms}\begin{eqnarray}
\langle\varphi^{m}\rangle_{0} & = & (uL^{d})^{-\frac{m}{4}}Y^{(m),\mathrm{mf}}(x_{\mathrm{mf}},y_{\mathrm{mf}}),\label{eq:phinscale}\\
\langle\left|\varphi\right|^{m}\rangle_{0} & = & (uL^{d})^{-\frac{m}{4}}Y_{\mathrm{abs}}^{(m),\mathrm{mf}}(x_{\mathrm{mf}},y_{\mathrm{mf}})\end{eqnarray}
\eS with the finite-size scaling functions\bS \label{eqs:ffabs(x,y)}\begin{eqnarray}
Y^{(m),\mathrm{mf}}(x,y) & = & \frac{\Xi_{m}(x,y)+(-1)^{m}\Xi_{m}(x,-y)}{\Xi_{0}(x,y)+\Xi_{0}(x,-y)},\label{eq:f(x,y)}\\
Y_{\mathrm{abs}}^{(m),\mathrm{mf}}(x,y) & = & \frac{\Xi_{m}(x,y)+\Xi_{m}(x,-y)}{\Xi_{0}(x,y)+\Xi_{0}(x,-y)}.\label{eq:fabs(x,y)}\end{eqnarray}
\eS Since, as discussed in Ref.~\cite{BrezinZinnJustin85}, the
zero mode $\varphi$ is related to the order parameter field $\phi(\vec{r})$
in real space via\begin{equation}
\varphi=\frac{1}{L^{d}}\sum_{j=1}^{L^{d}}\phi(\vec{r}_{j})\end{equation}
and consequently corresponds to the average order parameter per volume,
one immediately obtains the finite-size scaling forms of the quantities
defined in Sec.~\ref{sec:Finite-Size-Scaling-Analysis}, using Eqs.~(\ref{fssforms}).
Due to this correspondence, the magnetization $m(r,h;L)$ and the
susceptibility $\chi(r,h;L)$ are given by \bS \begin{eqnarray}
m(r,h;L) & = & \langle\varphi\rangle_{0}\label{eq:m1ss}\\
\beta^{-1}\chi(r,h;L) & = & L^{d}(\langle\varphi^{2}\rangle_{0}-\langle\varphi\rangle_{0}^{2})\label{eq:chi2aa}\end{eqnarray}
 \eS and therefore, according to Eq.~(\ref{eq:phinscale}), scale
as 

\bS \label{eqs:m_chi(r,h,L)}\begin{eqnarray}
m(r,h;L) & = & u^{-\frac{1}{4}}L^{-\frac{d}{4}}Y_{m}^{\mathrm{mf}}(x_{\mathrm{mf}},y_{\mathrm{mf}})\label{eq:m1}\\
\beta^{-1}\chi(r,h;L) & = & u^{-\frac{1}{2}}L^{\frac{d}{2}}Y_{\chi}^{\mathrm{mf}}(x_{\mathrm{mf}},y_{\mathrm{mf}})\label{eq:chi2}\end{eqnarray}
\eS with the zero mode finite-size scaling functions $Y_{m}^{\mathrm{mf}}(x,y)$
and $Y_{\chi}^{\mathrm{mf}}(x,y)$. The finite-size scaling forms
of the quantities\bS \begin{eqnarray}
m_{\mathrm{abs}}(r,h;L) & = & \langle|\varphi|\rangle_{0}\label{eq:mabs1ss}\\
\beta^{-1}\chi_{\mathrm{abs}}(r,h;L) & = & L^{d}(\langle\varphi^{2}\rangle_{0}-\langle|\varphi|\rangle_{0}^{2})\label{eq:chiabs2aa}\end{eqnarray}
 \eS are identical to Eqs.~(\ref{eqs:m_chi(r,h,L)}), respectively,
with the corresponding finite-size scaling functions $Y_{m_{\mathrm{abs}}}^{\mathrm{mf}}(x,y)$
and $Y_{\chi_{\mathrm{abs}}}^{\mathrm{mf}}(x,y)$. A further quantity
of interest is the dimensionless Binder cumulant\begin{equation}
U(r;L)=\left.1-\frac{\langle\varphi^{4}\rangle_{0}}{3\langle\varphi^{2}\rangle_{0}^{2}}\right|_{h=0}\label{eq:U(r,L)}\end{equation}
for which within the zero mode approximation one obtains the scaling
form\begin{equation}
U(r;L)=Y_{U}^{\mathrm{mf}}(x_{\mathrm{mf}})\end{equation}
 with the finite-size scaling function $Y_{U}^{\mathrm{mf}}(x)$. 

\begin{figure*}[t]
\begin{center}~\hfill{}\includegraphics[%
  clip,
  width=8cm,
  keepaspectratio]{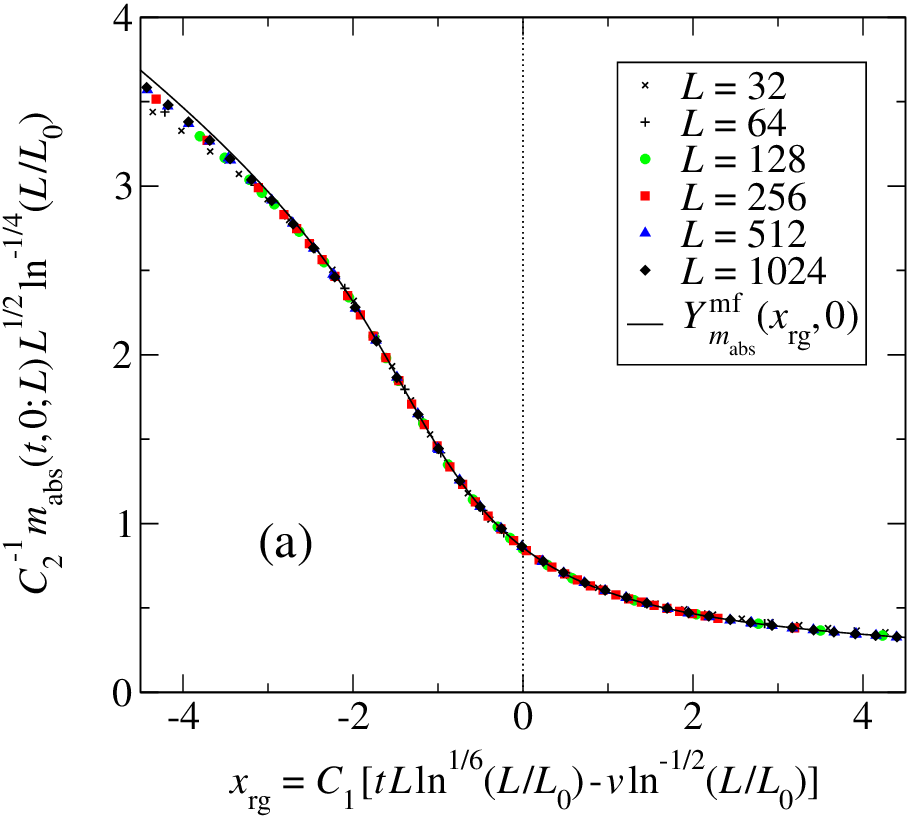}\hfill{}\includegraphics[%
  clip,
  width=8cm,
  keepaspectratio]{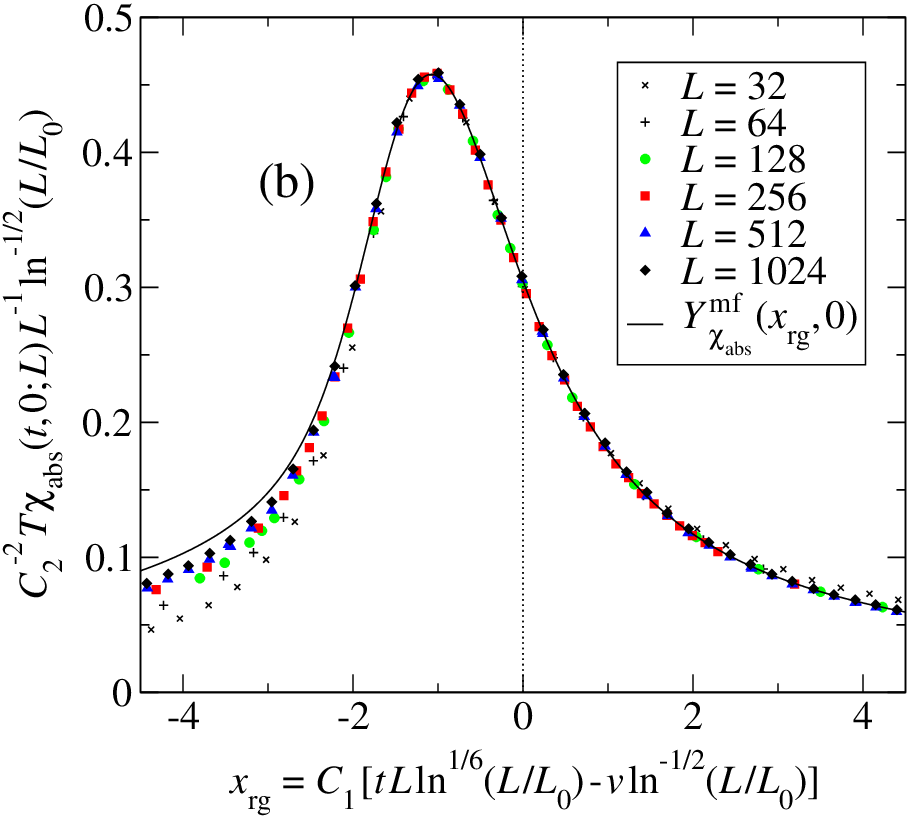}\hfill{}~ \\
~\hfill{}\includegraphics[%
  clip,
  width=8cm,
  keepaspectratio]{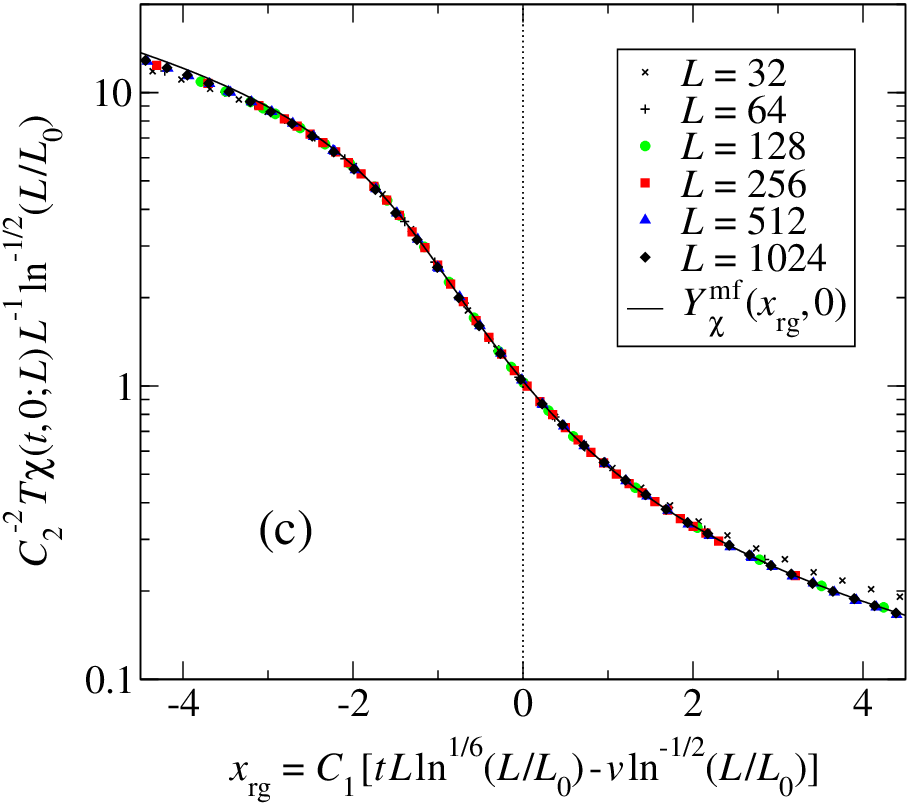}\hfill{}\includegraphics[%
  bb=6bp 4bp 443bp 397bp,
  clip,
  width=8cm,
  keepaspectratio]{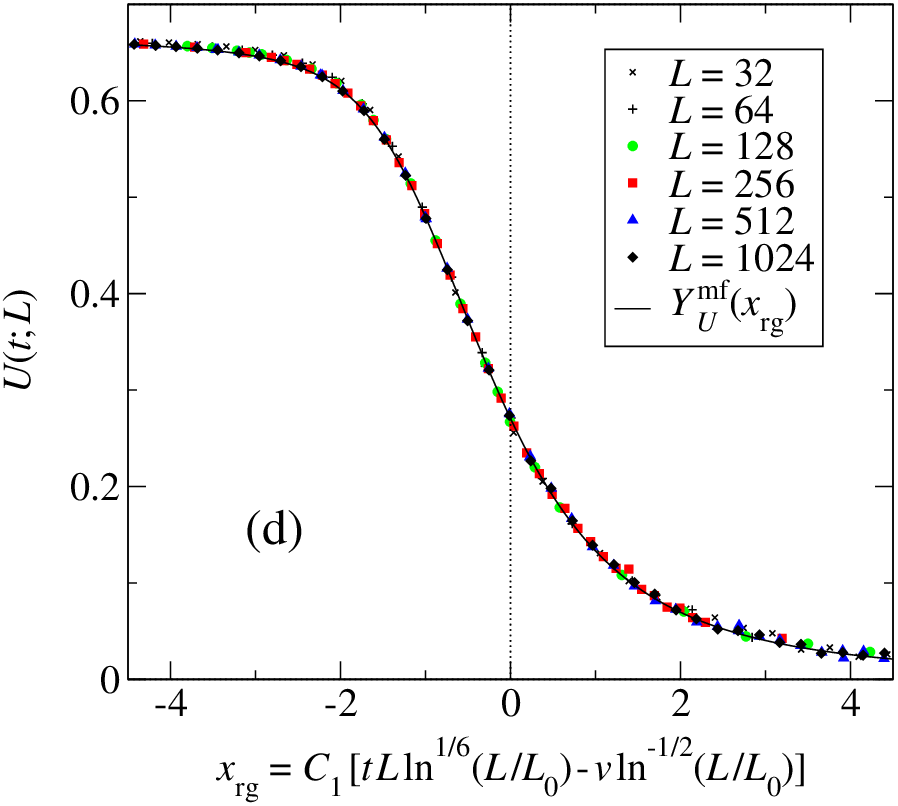}\hfill{}~\end{center}

\caption{Finite-size scaling plot of the magnetization $m_{\mathrm{abs}}(t,0;L)$
(a), the susceptibility $\chi_{\mathrm{abs}}(t,0;L)$ (b), the susceptibility
$\chi(t,0;L)$ (c), and the Binder cumulant $U(t;L)$ (d), versus
scaling argument $x_{\mathrm{rg}}$. The corresponding zero mode finite-size
scaling functions $Y_{i}^{\mathrm{mf}}(x,0)$ are displayed as solid
lines. Each data point was obtained by averaging over $10^{5}$ MCS.
\label{cap:data(t)}}
\end{figure*}
All zero mode finite-size scaling functions $Y_{i}^{\mathrm{mf}}(x,y)$
can be expressed as combinations of the functions $\Xi_{m}(x,y)$
defined in Eq.~(\ref{eq:Xi(x,y)}) with $m=\left\{ 0,1,2,4\right\} $.
Since to analyze the critical behavior, as it is done in the MC simulations,
either the temperature scaling variable or the field scaling variable
is kept at its critical point value, it is sufficient to evaluate
$\Xi_{m}(x,0)$ and $\Xi_{m}(0,y)$ for the pertinent values of $m$.
Analytical expressions for the needed functions can be found in the
Appendix.

\section{Monte Carlo Results \label{sec:Monte-Carlo-Results}}

\begin{figure*}[t]
\begin{center}~\hfill{}\includegraphics[%
  clip,
  width=8cm,
  keepaspectratio]{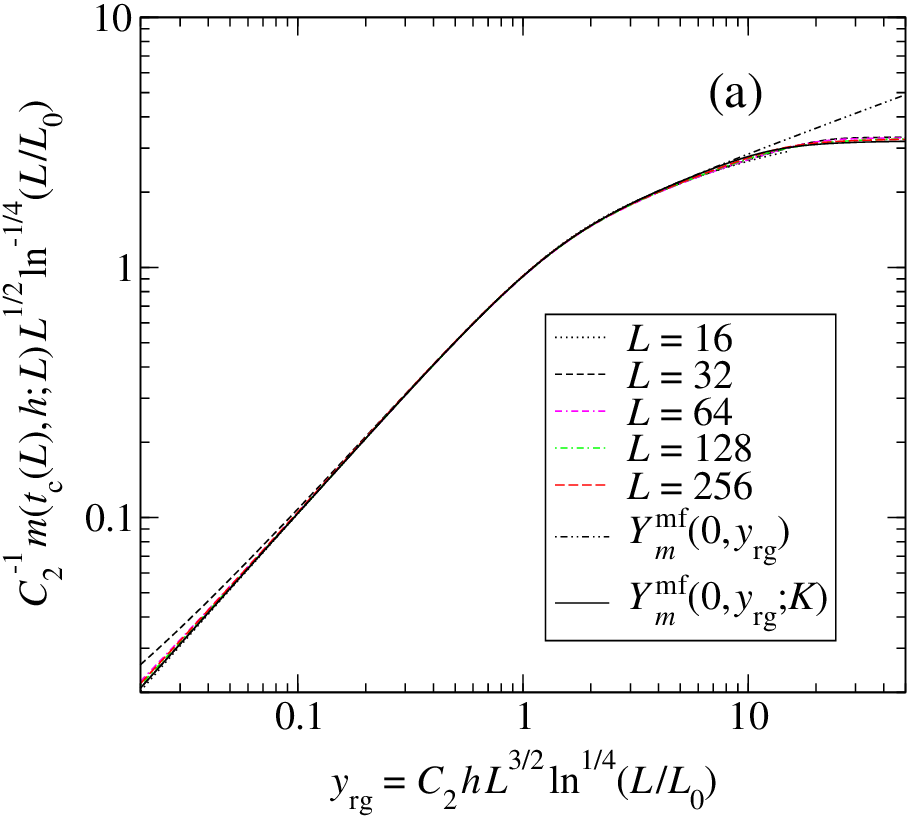}\hfill{}\includegraphics[%
  clip,
  width=8cm,
  keepaspectratio]{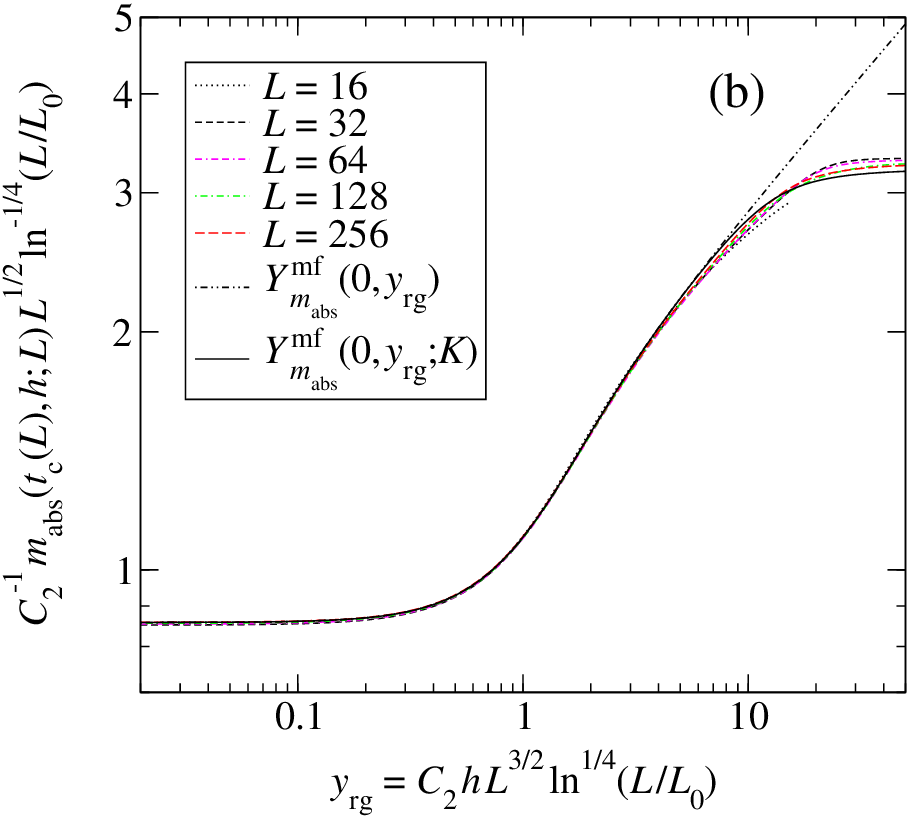}\hfill{}~\\
~\hfill{}\includegraphics[%
  bb=6bp 4bp 443bp 397bp,
  clip,
  width=8cm,
  keepaspectratio]{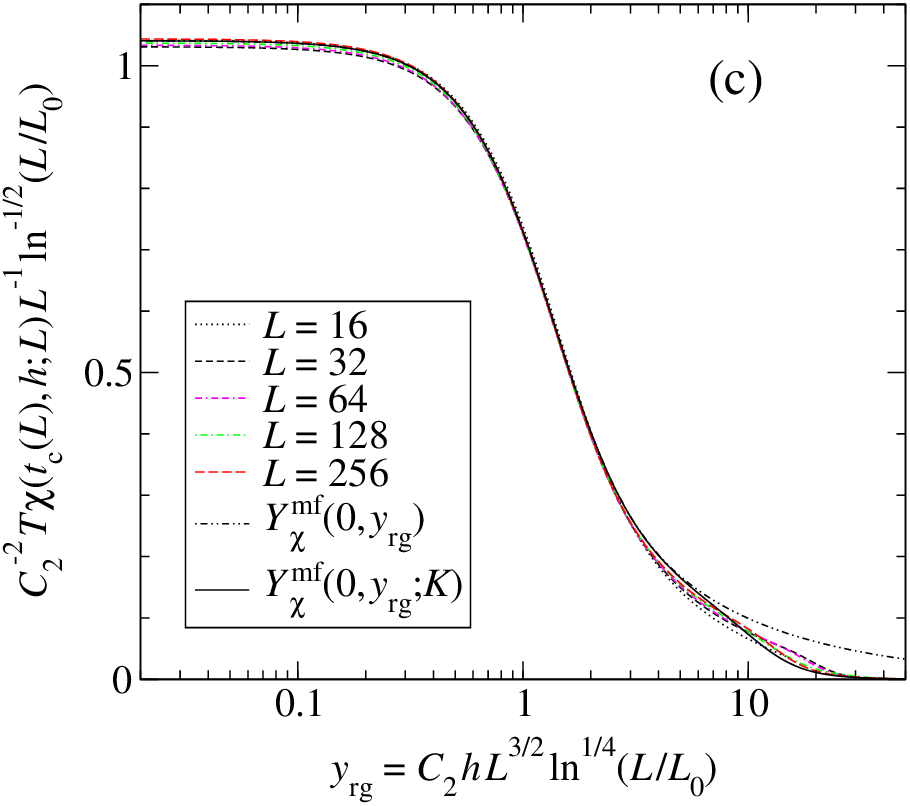}\hfill{}\includegraphics[%
  clip,
  width=8cm,
  keepaspectratio]{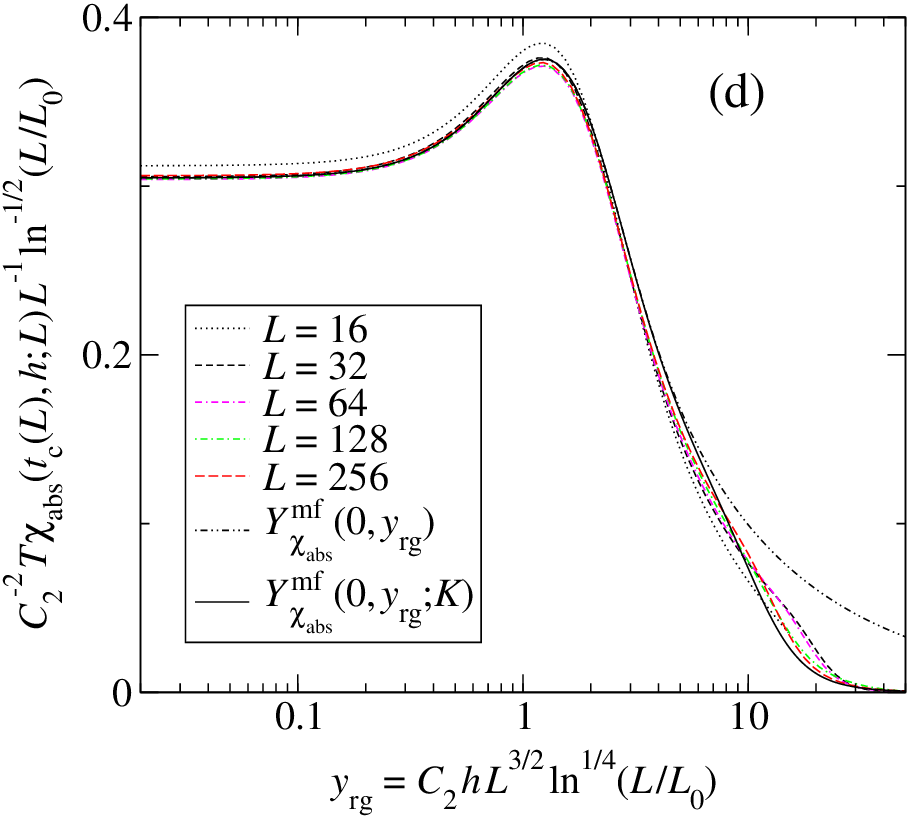}\hfill{}~\end{center}

\caption{Finite-size scaling plot of the magnetization $m(t_{c}(L),h;L)$
(a) and $m_{\mathrm{abs}}(t_{c}(L),h;L)$ (b), and the susceptibility
$\chi(t_{c}(L),h;L)$ (c) and $\chi_{\mathrm{abs}}(t_{c}(L),h;L)$
(d), versus scaling argument $y_{\mathrm{rg}}$. The size of the histograms
that were used to obtain the curves amounts $K=2\times10^{6}$ MCS
for each system length $L$. The zero mode finite-size scaling functions
$Y_{i}^{\mathrm{mf}}(x,y;K)$ are plotted as solid lines where the
cutoff parameter takes the value $q_{\Lambda}(2\times10^{6})\simeq3.2201$
(see text). \label{cap:data(h)}}
\end{figure*}
In this section we present the results of MC simulations that were
carried out for the model {[}Eq.~(\ref{eqs:Model}){]} with $(\omega_{\para},\omega_{\bot},J)=(1,1,0)$.
We have used the Wolff cluster algorithm \cite{Wolff89} for long-range
spin models proposed by Luijten and Blöte \cite{LuijtenBloete95}.
To study the properties of the model in the presence of an external
magnetic field, a histogram reweighting technique \cite{Ferrenberg88}
was used. 

In the simulations quadratic spin systems with $L:=L_{\para}=L_{\bot}=\{32,64,128,256,512,1024\}$
were considered and we started with recording the magnetization $m_{\mathrm{abs}}$,
the susceptibilities $\chi$ and $\chi_{\mathrm{abs}}$, and the Binder
cumulant $U$ at zero field for various temperatures close to the
corresponding mean-field critical temperature $T_{\mathrm{c}}^{\mathrm{mf}}$.
This temperature is given by $T_{\mathrm{c}}^{\mathrm{mf}}=\widetilde{J}(\vec{0})$
(see, \eg, Ref.~\cite{Aharony82}) where $\widetilde{J}(\vec{k})$
denotes the Fourier transform of the pair coupling Eq.~(\ref{Interaction}).
Setting the coupling constant to $J=0$ the evaluation of this expression
yields, the divergent term at $\vec{r}=\vec{0}$ is excluded,\begin{equation}
T_{\mathrm{c}}^{\mathrm{mf}}=\sideset{}'\sum_{\mathbf{r}\in\mathbb{Z}^{2}}\frac{\omega_{\parallel}^{{}}r_{\parallel}^{2}+\omega_{\bot}^{{}}r_{\bot}^{2}}{\left|\mathbf{r}\right|^{5}}=\frac{1}{2}(\omega_{\parallel}+\omega_{\bot})\Theta({\textstyle \frac{3}{2}})\end{equation}
with the two-dimensional lattice sum~\cite{Hucht95}\begin{equation}
\Theta(s)=4^{1-s}\zeta(s)\left(\zeta(s,{\textstyle \frac{1}{4}})-\zeta(s,{\textstyle \frac{3}{4}})\right),\label{eq:Theta(s)}\end{equation}
where $\zeta(s,a)$ denotes the generalized Riemann Zeta function.
Setting $\omega_{\para}=\omega_{\bot}=1$, the mean-field critical
temperature takes the value\begin{equation}
T_{\mathrm{c}}^{\mathrm{mf}}=\Theta({\textstyle \frac{3}{2}})\simeq9.0336.\end{equation}
 According to the finite-size scaling relations listed in Sec.~\ref{sec:Finite-Size-Scaling-Analysis}
the UFSS functions were evaluated from the MC data and plotted against
the temperature scaling variable $x_{\mathrm{rg}}$ (see Fig.~\ref{cap:data(t)}).
The data collapse was achieved by adjusting the critical temperature
$T_{\mathrm{c}}\equiv T_{\mathrm{c}}(\infty)$ and the parameters
$v$ and $L_{0}$ in the following way. First we determined $L_{0}$
from the requirement that the maximum of the scaled susceptibility
$T\chi_{\mathrm{abs}}(t,0;L)L^{-1}\ln^{-\frac{1}{2}}({\textstyle \frac{L}{L_{0}}})$
collapses for different $L$ {[}Fig.~\ref{cap:data(t)}b{]}, as this
peak height is independent of $T_{\mathrm{c}}$ and $v$. After that
we adjusted $T_{\mathrm{c}}$ and $v$ until the scaled cumulant $U(t;L)$
vs. $x_{\mathrm{rg}}$ {[}Eqs.~(\ref{eq:U(t,L)},\ref{eq:xrg}){]}
collapses {[}Fig.~\ref{cap:data(t)}d{]} and fits the well known
critical value \cite{BrezinZinnJustin85,LuijtenBloete97}\begin{equation}
Y_{U}(0)=1-\frac{\Gamma(\frac{1}{4})^{4}}{24\pi^{2}}=0.27052\ldots\end{equation}
 at $x_{\mathrm{rg}}{=}0$. Finally, in order to compare the numerical
data to the zero mode finite-size scaling functions listed in Sec.~\ref{sec:Mean-Field-Theory},
these functions were fitted to the numerical data by tuning the nonuniversal
metric factors $C_{1}$ and $C_{2}$. The values of all parameters
as they were determined from this analysis are listed in Table~\ref{Tab:Parameterfss}.
The whole data analysis was done using \noun{Fsscale} \cite{fsscale}.

\begin{table}[b]
\begin{center}\begin{tabular}{|c|c|c|c|c|}
\hline 
$T_{\mathrm{c}}(\infty)$&
$L_{0}$&
$v$&
$C_{1}$&
$C_{2}$\tabularnewline
\hline
\hline 
$8.0302(3)$&
$3.0(2)$&
$1.16(2)$&
$0.735(10)$&
$0.92(1)$\tabularnewline
\hline
\end{tabular}\end{center}

\caption{Values of the fit parameters used throughout this work.\label{Tab:Parameterfss}}
\end{table}
In addition to the temperature dependence we also studied the dependence
of the quantities $m$, $m_{\mathrm{abs}}$, $\chi$, and $\chi_{\mathrm{abs}}$
on an external field at the temperature\begin{equation}
T_{\mathrm{c}}(L)=T_{\mathrm{c}}(\infty)(1+vL^{-1}\ln^{-\frac{2}{3}}({\textstyle \frac{L}{L_{0}}}))\end{equation}
 that can be regarded as an effective critical temperature of the
finite system (see, \eg, Ref.~\cite{BinNauPrivYoung85}). Note that
$t_{\mathrm{c}}(L)=T_{\mathrm{c}}(L)/T_{\mathrm{c}}(\infty)-1$ corresponds
to the value of $t$ for which the shifted reduced temperature $\hat{t}$
as defined in Eq.~(\ref{eq:dfdffd}) and consequently the temperature
scaling variable $x_{\mathrm{rg}}$ {[}Eq.~(\ref{eq:xrg}){]} vanishes,
as $\hat{t}=t-t_{\mathrm{c}}(L)$. Due to this choice the numerical
data obtained at this temperature and nonzero fields might then be
compared to the corresponding zero mode finite-size scaling functions
$Y_{i}^{\mathrm{mf}}(0,y)$ within a finite-size scaling plot. Therefore
we stored for each of the system lengths $L=\{16,32,64,128,256\}$
a magnetization histogram at the corresponding temperature $T_{\mathrm{c}}(L)$
and zero external field, \ie at $x_{\mathrm{rg}}=y_{\mathrm{rg}}=0$.
Using these histograms, the considered quantities were extrapolated
to nonzero fields \cite{Ferrenberg88} and plotted as implied by their
finite-size scaling forms (see Fig.~\ref{cap:data(h)}). The values
of the fit parameters required for these plots are taken consistently
as they were determined from the temperature runs (cf.~Table~\ref{Tab:Parameterfss}).
Since the reweighting technique allows the extrapolation of the quantities
to arbitrary values of the external field, the data are displayed
as continuous lines. 

Within the intervals of the scaling variables $x_{\mathrm{rg}}$ and
$y_{\mathrm{rg}}$ that were considered in the simulations we find
excellent agreement of the MC data with the zero mode theory. In Figs.~\ref{cap:data(t)}(a,b)
merely for negative values of the temperature scaling variable $x_{\mathrm{rg}}$
remarkable deviations occur and the MC data with increasing system
size $L$ converge to the corresponding zero mode finite-size scaling
function. This effect indicates the presence of further corrections
that are expected to vanish in the limit $L\rightarrow\infty$ (see
also Ref.~\cite{LuiBinBloe99} that refers to the five-dimensional
short-range nearest neighbor Ising model). 

\begin{figure}[tb]
\begin{center}\includegraphics[%
  clip,
  width=8cm,
  keepaspectratio]{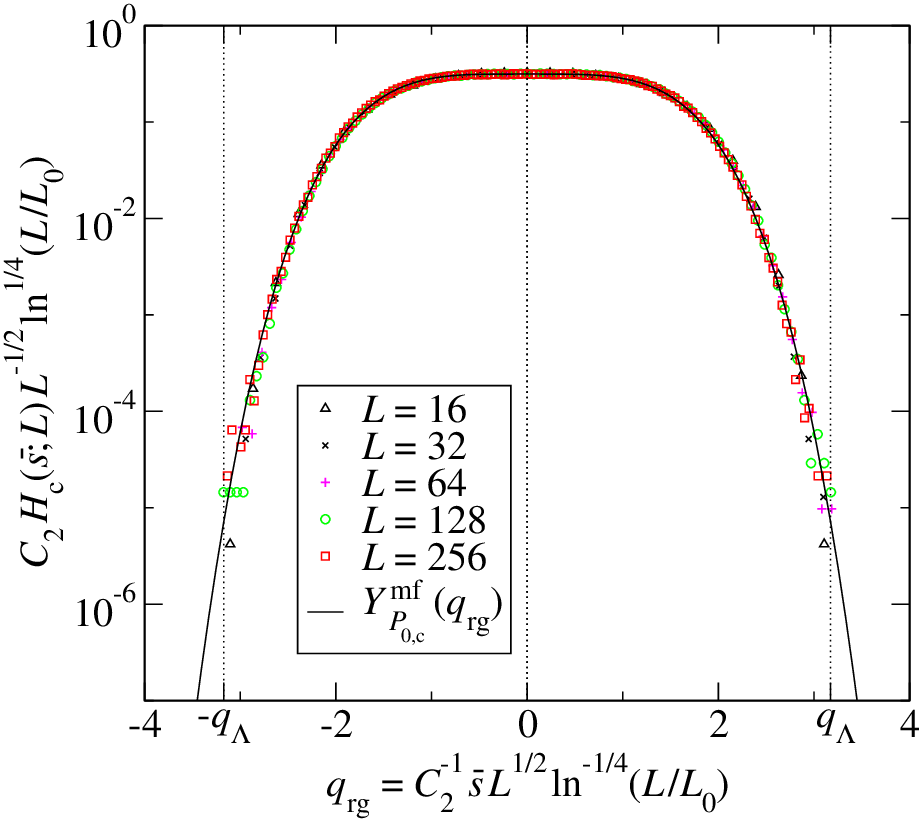}\end{center}

\caption{Finite-size scaling plot of the critical histogram $H_{\mathrm{c}}(\bar{s};L)$.
The size of the histogram depicted here amounts $K=10^{6}$ MCS for
each system length $L$. According to Eq.~(\ref{eq:cutoff-def})
this corresponds to a  cutoff value of $q_{\Lambda}(10^{6})\simeq3.1728$.
The value of the fit parameter $L_{0}$ and the nonuniversal metric
factor $C_{2}$ are taken as they are listed in Table~\ref{Tab:Parameterfss}.
\label{cap:cutoff}}
\end{figure}
\begin{figure*}[t]
\begin{center}~\hfill{}\includegraphics[%
  clip,
  width=8cm,
  keepaspectratio]{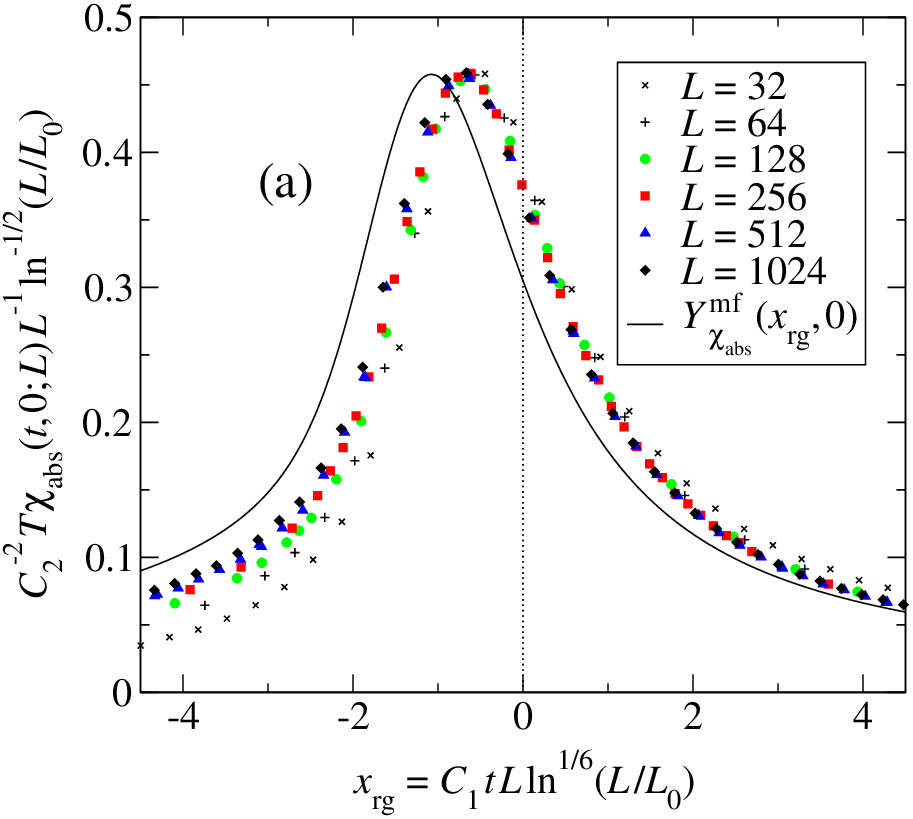}\hfill{}\includegraphics[%
  clip,
  width=8cm,
  keepaspectratio]{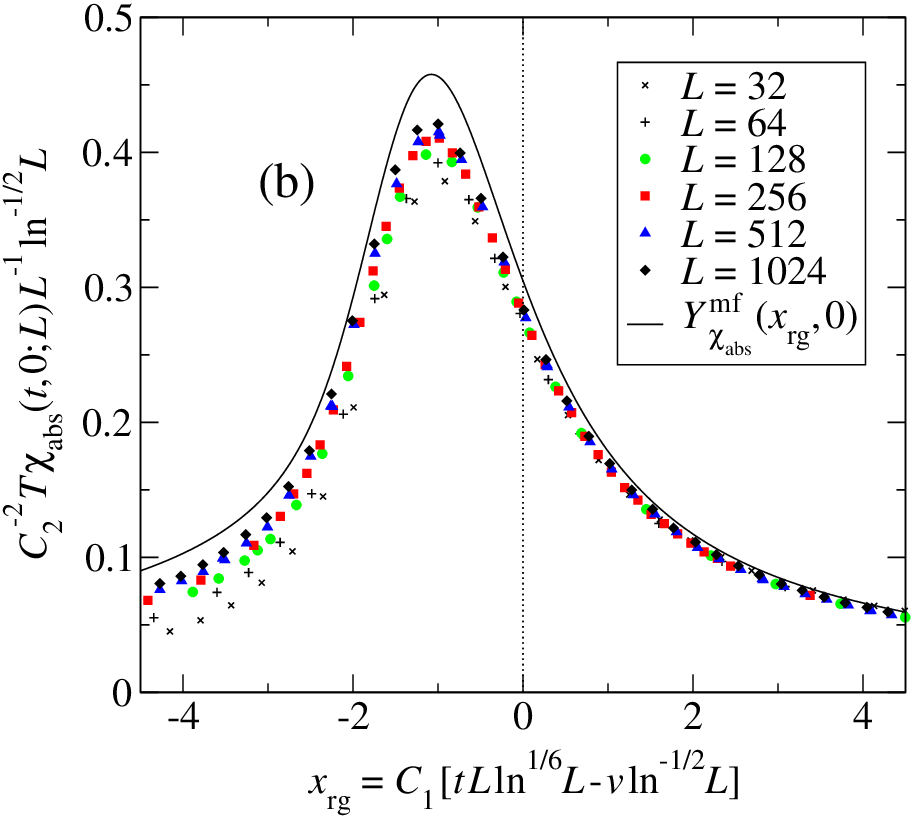}\hfill{}~\end{center}

\caption{Finite-size scaling plot of the susceptibility $\chi_{\mathrm{abs}}(t,0;L)$
without the shift correction ($v=0$) (a) and with $L_{0}=1$ (b).
\label{cap:chiabs_vL(t)}}
\end{figure*}
The deviations visible in Fig.~\ref{cap:data(h)} for large values
of the field scaling variable $y_{\mathrm{rg}}$ are due to the finite
size of the histograms that were used to obtain the curves shown there
by means of the reweighting technique. They originate from the fact
that a magnetization histogram of size $K$ that is based on $K$
values of the average spin $\bar{s}$ {[}Eq.~(\ref{eq:avgspin}){]}
sampled in a MC simulation, in the mean does not contain values of
$\bar{s}$ occurring in a simulation with a lower probability than
$1/K$. As will be demonstrated in the following, these deviations
can also be reproduced within the zero mode theory by truncating the
zero mode probability distribution $\mathcal{P}_{0}(\varphi)$, Eq.~(\ref{zmaGibbsdistr}),
from which we start and perform the rescaling, Eq.~(\ref{eq:rescaling}).
After that we set $x_{\mathrm{mf}}=0$ and $y_{\mathrm{mf}}=0$ since
the magnetization histograms we used to obtain the curves shown in
Fig.~\ref{cap:data(h)} were recorded in the simulations also for
vanishing $x_{\mathrm{rg}}$ and $y_{\mathrm{rg}}$. The corresponding
normalized zero mode probability distribution denoted as $\mathcal{P}_{0,\mathrm{c}}(\varphi)$
then reads\bS{}\label{Probab-Hist}\begin{equation}
\mathcal{P}_{0,\mathrm{c}}(\varphi)=(uL^{d})^{\frac{1}{4}}Y_{\mathcal{P}_{0,\mathrm{c}}}^{\mathrm{mf}}((uL^{d})^{\frac{1}{4}}\varphi)\end{equation}
with the zero mode finite-size scaling function\begin{equation}
Y_{\mathcal{P}_{0,\mathrm{c}}}^{\mathrm{mf}}(q)=\frac{3^{\frac{3}{4}}}{2\Gamma({\textstyle \frac{1}{4}})}e^{-\frac{27}{256}q^{4}}\label{eq:zero-mode-Y}\end{equation}
\eS{}and hence decreases monotonically for increasing $|\varphi|$.
So to reproduce the effect that is due to the finite size of the histograms
we chop the tails of this distribution by limiting the integration
range in Eq.~(\ref{eq:avgzma}) and the zero mode partition function,
Eq.~(\ref{eq:ZMPF}), to a finite interval $[-\varphi_{\Lambda},\varphi_{\Lambda}]$.
To make an appropriate definition of the cut-off parameter $\varphi_{\Lambda}\equiv\varphi_{\Lambda}(K)>0$
we use an estimation that is known from extreme value statistics \cite{KotzNadarajah00}.
We assume that in the mean only one out of $K$ measured values of
the average order parameter $\varphi$ lies outside the interval $]-\varphi_{\Lambda},\varphi_{\Lambda}[$,
\ie it fulfills $|\varphi|\geq\varphi_{\Lambda}$. This implies\begin{equation}
P(|\varphi|\geq\varphi_{\Lambda})=2\int_{\varphi_{\Lambda}}^{\infty}\mathrm{d}\varphi\,\mathcal{P}_{0,\mathrm{c}}(\varphi)\stackrel{!}{=}\frac{1}{K}\end{equation}
from which the cutoff $\varphi_{\Lambda}$ is implicitly defined.
Replacing $\mathcal{P}_{0,\mathrm{c}}(\varphi)$ according to Eqs.~(\ref{Probab-Hist})
the evaluation of the resulting integral \noun{}yields\begin{equation}
2\int_{q_{\Lambda}}^{\infty}\mathrm{d}q\, Y_{\mathcal{P}_{0,\mathrm{c}}}^{\mathrm{mf}}(q)=\frac{\Gamma({\textstyle \frac{1}{4}},{\textstyle \frac{27}{256}}q_{\Lambda}^{4})}{\Gamma({\textstyle \frac{1}{4}})}\stackrel{!}{=}\frac{1}{K},\label{eq:cutoff-def}\end{equation}
where $q_{\Lambda}=(uL^{d})^{\frac{1}{4}}\varphi_{\Lambda}$ and $\Gamma(a,b)$
denotes the incomplete Gamma function. Expanding this expression and
separating $q_{\Lambda}\equiv q_{\Lambda}(K)$ yields to leading order
a logarithmic dependence on the histogram size $K$ of the form\begin{equation}
q_{\Lambda}(K)\stackrel{K\rightarrow\infty}{\sim}\sqrt{\frac{8}{3}}\!\left(\frac{4}{3}\ln({\textstyle \frac{K}{K_{0}}})-\ln\!\left(\frac{4}{3}\ln({\textstyle \frac{K}{K_{0}}})\right)\right)^{\frac{1}{4}}\end{equation}
with the constant $K_{0}=(3/4)^{\frac{3}{4}}\Gamma({\textstyle \frac{1}{4}}).$

To demonstrate the validity of the definition of the cutoff $\varphi_{\Lambda}$,
respectively, $q_{\Lambda}$, Fig.~\ref{cap:cutoff} shows a finite-size
scaling plot of the critical magnetization histogram $H_{\mathrm{c}}$
that was recorded for $x_{\mathrm{rg}}=y_{\mathrm{rg}}=0$. Starting
from the finite-size scaling form of the magnetization, Eq.~(\ref{eq:m(t,h,L)}),
it is straightforward to show that this histogram scales as\begin{equation}
H_{\mathrm{c}}(\bar{s};L)\sim C_{2}^{-1}L^{\frac{1}{2}}\ln^{-\frac{1}{4}}({\textstyle \frac{L}{L_{0}}})Y_{H_{\mathrm{c}}}(q_{\mathrm{rg}})\end{equation}
with the finite-size scaling function $Y_{H_{\mathrm{c}}}(q)$ and
the scaling variable $q_{\mathrm{rg}}=C_{2}^{-1}\bar{s}L^{\frac{1}{2}}\ln^{-\frac{1}{4}}({\textstyle \frac{L}{L_{0}}})$.
As is expected also in Fig.~\ref{cap:cutoff}, the MC data are found
to be in very good agreement with the zero mode expression $Y_{\mathcal{P}_{0,\mathrm{c}}}^{\mathrm{mf}}(q)$
{[}Eq.~(\ref{eq:zero-mode-Y}){]} of the finite-size scaling function
$Y_{H_{\mathrm{c}}}(q)$. It furthermore turns out that all data points
lie accurately within the interval $[-q_{\Lambda}(K),q_{\Lambda}(K)]$
that corresponds to the size of the displayed histogram. The zero
mode finite-size scaling functions $Y_{i}^{\mathrm{mf}}(x,y;K)$ that
were evaluated using the truncated zero mode probability distribution
with the cutoff $q_{\Lambda}(K)$ are also shown in Fig.~\ref{cap:data(h)},
 and nicely agree with the extrapolated data. 

Before finishing this section we also want to discuss the importance
of the corrections to the leading orders in the expansions in Eqs.~(\ref{eq:expansions})
that are determined by the parameters $L_{0}$ and $v$. For that
purpose Figs.~\ref{cap:chiabs_vL(t)}(a,b) show the finite-size scaling
plot of the susceptibility $\chi_{\mathrm{abs}}(t,0;L)$ versus scaling
variable $x_{\mathrm{rg}}$ (Fig.~\ref{cap:data(t)}(b)) where we
set $v=0$ in Fig.~\ref{cap:chiabs_vL(t)}(a) and $L_{0}=1$ in Fig.~\ref{cap:chiabs_vL(t)}(b).
The values of the other parameters entering the plots are taken from
Table~\ref{Tab:Parameterfss}. As is expected, neglecting one of
the corrections causes a significant displacement between the MC data
and the zero mode finite-size scaling function $Y_{\chi_{\mathrm{abs}}}^{\mathrm{mf}}(x_{\mathrm{rg}},0)$.
This effect could partially be compensated by readjusting the remaining
fit parameters, but this procedure would in the case of Fig.~\ref{cap:chiabs_vL(t)}(a)
lead to a wrong determination of the critical temperature $T_{\mathrm{c}}(\infty)$
(see also discussions in Ref.~\cite{LuijtenBloete96} that refer
to spin models above the upper critical dimensionality).

\section{Conclusions}

We have introduced a two-dimensional long-range spin model that displays
both isotropic and anisotropic phase transitions and, in particular,
strongly anisotropic phase transitions. As a first stage the critical
behavior of the model in the isotropic case for which it is found
to be at its upper critical dimensionality was investigated. For that
purpose we have carried out Monte Carlo simulations and studied the
temperature and field dependence of several quantities. Using results
of the renormalization group, the numerical data obtained for different
system sizes were analyzed by means of a finite-size scaling analysis.
It turns out that beside a size-dependent shift that has already been
discussed in the literature a characteristic length $L_{0}$ that
was inserted to the logarithms is an important correction that must
not be neglected. 

Furthermore, the collapsed data were compared to the zero mode (mean-field)
theory and found to be in excellent agreement. It turns out that the
logarithmic corrections typically occurring at the upper critical
dimensionality do only enter the finite-size scaling functions through
their arguments, whereby these functions were derived from zero mode
theory. This shows that at least in the present case the concept of
universal finite-size scaling functions can be extended to the upper
critical dimensionality. 

Finally we note that the numerical results strongly indicate the validity
of the zero mode theory at the upper critical dimensionality and might
shed new light on recent controversial discussions about its correctness
for $d\geq d_{\mathrm{u}}$ \cite{ChenDohm98a,LuiBinBloe99}. 

As it will be subject of a future work it is desirable to extend the
analysis shown above to the anisotropic case $\omega_{\parallel}\neq\omega_{\perp}$.
In particular, the critical behavior of the model should be investigated
when approaching the strongly anisotropic cases $\omega_{\parallel}=-2\omega_{\perp}$
and $-2\omega_{\parallel}=\omega_{\perp}$, respectively. 

\begin{acknowledgments}
The authors would like to thank S.~Lübeck, Professor K.~D.~Usa\-del
and Professor H.~W.~Diehl for valuable discussions.
\end{acknowledgments}
\appendix

\section{The Functions $\mathbf{\Xi_{m}\mathrm{(x,0)}}$ And $\mathbf{\Xi_{m}\mathrm{(0,y)}}$
\label{sec:The-Functions-Xi}}

The evaluation of the function $\Xi_{m}\mathrm{(x,0)}$ with \noun{Mathematica~\cite{MMA4}}
yields\begin{eqnarray}
\Xi_{m}(x,0) & = & 3^{-\frac{3}{4}(m+1)}4^{m}\!\left(\Gamma\!\left({\textstyle \frac{m+1}{4}}\right)\,_{1}F_{1}\!\left({\textstyle \frac{m+1}{4};\frac{1}{2};x^{2}}\right)\right.\nonumber \\
 &  & \left.-2x\Gamma\!\left({\textstyle \frac{m+3}{4}}\right)\,_{1}F_{1}\!\left({\textstyle \frac{m+3}{4};\frac{3}{2};x^{2}}\right)\right)\label{eq:Xim(x,0)}\end{eqnarray}
with the confluent hypergeometric function $_{1}F_{1}(a;b;x)$. This
expression is valid for $x\in\mathbb{R}$ and $m\in\mathbb{R}^{>-1}$
and can be simplified further for a given integer value of the parameter
$m$. For $m=\{0,1,2,4\}$ one obtains \bS\begin{eqnarray}
\Xi_{0}(x,0) & = & 3^{-\frac{3}{4}}e^{\frac{x^{2}}{2}}\Upsilon_{\frac{1}{4}}(x),\label{eq:Xi0(x,0)}\\
\Xi_{1}(x,0) & = & \frac{4}{3}\sqrt{\frac{\pi}{3}}e^{x^{2}}\erfc(x),\label{eq:Xi1(x,0)}\\
\Xi_{2}(x,0) & = & \frac{8}{9}3^{-\frac{1}{4}}e^{\frac{x^{2}}{2}}\!\left(\Upsilon_{\frac{3}{4}}(x)-x\Upsilon_{\frac{1}{4}}(x)\right),\label{eq:Xi2(x,0)}\\
\Xi_{4}(x,0) & = & \frac{64}{81}3^{\frac{1}{4}}e^{\frac{x^{2}}{2}}\!\left(2x^{2}\Upsilon_{\frac{1}{4}}(x)-3x\Upsilon_{\frac{3}{4}}(x)-\Upsilon_{\frac{5}{4}}(x)\right),\nonumber \\
\label{eq:Xi4(x,0)}\end{eqnarray}
\eS where the function\begin{equation}
\Upsilon_{a}(x)=\pi(x^{2})^{a}\left(I_{-a}\!\left({\textstyle \frac{1}{2}}x^{2}\right)-\sgn(x)I_{a}\!\left({\textstyle \frac{1}{2}}x^{2}\right)\right)\label{eq:Upsa(x)}\end{equation}
with the modified Bessel function of the first kind $I_{a}(x)$ has
been introduced, and $\erfc(x)$ denotes the complementary error function.
The function $\Upsilon_{a}(x)$ is well behaved through zero argument
for the pertinent noninteger positive values of $a$.  Assuming $x>0$,
Eq.~(\ref{eq:Upsa(x)}) simplifies to\begin{equation}
\Upsilon_{a}(x)\stackrel{x>0}{=}2x^{2a}K_{a}\!\left({\textstyle \frac{1}{2}}x^{2}\right)\sin(\pi a)\end{equation}
with the modified Bessel function of the second kind $K_{a}(x)$. 

An analogous treatment of the function $\Xi_{m}\mathrm{(0,y)}$ results
in\begin{eqnarray}
\Xi_{m}(0,y) & = & \frac{1}{4}\sum_{k=1}^{4}\left({\textstyle \frac{256}{27}}\right)^{\!\!\frac{m+k}{4}}\frac{y^{k-1}}{\Gamma(k)}\Gamma\!\left({\textstyle \frac{m+k}{4}}\right)\label{eq:Xi(0,y)}\\
 &  & {}\times{_{2}F_{4}}\!\left({\textstyle \frac{m+k}{4},1;\frac{k}{4},\frac{k+1}{4},\frac{k+2}{4},\frac{k+3}{4};\frac{y^{4}}{27}}\right)\nonumber \end{eqnarray}
where $_{p}F_{q}(a_{1},\ldots,a_{p};b_{1},\ldots,b_{q};x)$ denotes
the generalized hypergeometric function. This expression is also valid
for $y\in\mathbb{R}$ and $m\in\mathbb{R}^{>-1}$ and cannot be simplified
further for a given value of $m$ involving less general functions.
Finally let us note that the function $\Xi_{m}(x,y)$ fulfills the
recursion relations \bS\begin{eqnarray}
\Xi_{m+2}(x,y) & = & -\frac{8}{3\sqrt{3}}\frac{\partial}{\partial x}\Xi_{m}(x,y),\\
\Xi_{m+1}(x,y) & = & \frac{\partial}{\partial y}\Xi_{m}(x,y)\end{eqnarray}
 \eS as follows from Eq.~(\ref{eq:Xi(x,y)}).

\bibliographystyle{apsrev}
\bibliography{Physik}

\begin{thebibliography}{33}
\expandafter\ifx\csname natexlab\endcsname\relax\def\natexlab#1{#1}\fi
\expandafter\ifx\csname bibnamefont\endcsname\relax
  \def\bibnamefont#1{#1}\fi
\expandafter\ifx\csname bibfnamefont\endcsname\relax
  \def\bibfnamefont#1{#1}\fi
\expandafter\ifx\csname citenamefont\endcsname\relax
  \def\citenamefont#1{#1}\fi
\expandafter\ifx\csname url\endcsname\relax
  \def\url#1{\texttt{#1}}\fi
\expandafter\ifx\csname urlprefix\endcsname\relax\def\urlprefix{URL }\fi
\providecommand{\bibinfo}[2]{#2}
\providecommand{\eprint}[2][]{\url{#2}}

\bibitem[{\citenamefont{Luijten and Blöte}(1995)}]{LuijtenBloete95}
\bibinfo{author}{\bibfnamefont{E.}~\bibnamefont{Luijten}} \bibnamefont{and}
  \bibinfo{author}{\bibfnamefont{H.~W.~J.} \bibnamefont{Blöte}},
  \bibinfo{journal}{Int. J.~Mod. Phys.~C} \textbf{\bibinfo{volume}{6}},
  \bibinfo{pages}{359} (\bibinfo{year}{1995}).

\bibitem[{\citenamefont{Luijten and Blöte}(1997)}]{LuijtenBloete97}
\bibinfo{author}{\bibfnamefont{E.}~\bibnamefont{Luijten}} \bibnamefont{and}
  \bibinfo{author}{\bibfnamefont{H.~W.~J.} \bibnamefont{Blöte}},
  \bibinfo{journal}{Phys. Rev.~B} \textbf{\bibinfo{volume}{56}},
  \bibinfo{pages}{8945} (\bibinfo{year}{1997}).

\bibitem[{\citenamefont{Luijten et~al.}(1996)\citenamefont{Luijten, Blöte, and
  Binder}}]{LuiBloeBin96}
\bibinfo{author}{\bibfnamefont{E.}~\bibnamefont{Luijten}},
  \bibinfo{author}{\bibfnamefont{H.~W.~J.} \bibnamefont{Blöte}},
  \bibnamefont{and} \bibinfo{author}{\bibfnamefont{K.}~\bibnamefont{Binder}},
  \bibinfo{journal}{Phys. Rev.~E} \textbf{\bibinfo{volume}{54}},
  \bibinfo{pages}{4626} (\bibinfo{year}{1996}).

\bibitem[{\citenamefont{Luijten et~al.}(1997)\citenamefont{Luijten, Blöte, and
  Binder}}]{LuiBloeBin97}
\bibinfo{author}{\bibfnamefont{E.}~\bibnamefont{Luijten}},
  \bibinfo{author}{\bibfnamefont{H.~W.~J.} \bibnamefont{Blöte}},
  \bibnamefont{and} \bibinfo{author}{\bibfnamefont{K.}~\bibnamefont{Binder}},
  \bibinfo{journal}{Phys. Rev.~E} \textbf{\bibinfo{volume}{56}},
  \bibinfo{pages}{6540} (\bibinfo{year}{1997}).

\bibitem[{\citenamefont{Hucht}(2002)}]{Hucht02a}
\bibinfo{author}{\bibfnamefont{A.}~\bibnamefont{Hucht}},
  \bibinfo{journal}{J.~Phys A: Math. Gen.} \textbf{\bibinfo{volume}{35}},
  \bibinfo{pages}{L481} (\bibinfo{year}{2002}).

\bibitem[{\citenamefont{Hornreich et~al.}(1975)\citenamefont{Hornreich, Luban,
  and Shtrikman}}]{Hornreich75}
\bibinfo{author}{\bibfnamefont{R.~M.} \bibnamefont{Hornreich}},
  \bibinfo{author}{\bibfnamefont{M.}~\bibnamefont{Luban}}, \bibnamefont{and}
  \bibinfo{author}{\bibfnamefont{S.}~\bibnamefont{Shtrikman}},
  \bibinfo{journal}{Phys. Rev. Lett.} \textbf{\bibinfo{volume}{35}},
  \bibinfo{pages}{1678} (\bibinfo{year}{1975}).

\bibitem[{\citenamefont{Selke}(1992)}]{Selke92}
\bibinfo{author}{\bibfnamefont{W.}~\bibnamefont{Selke}}, in
  \emph{\bibinfo{booktitle}{Phase Transitions and Critical Phenomena}}, edited
  by \bibinfo{editor}{\bibfnamefont{C.}~\bibnamefont{Domb}} \bibnamefont{and}
  \bibinfo{editor}{\bibfnamefont{J.~L.} \bibnamefont{Lebowitz}}
  (\bibinfo{publisher}{Academic Press}, \bibinfo{address}{London},
  \bibinfo{year}{1992}), vol.~\bibinfo{volume}{15}.

\bibitem[{\citenamefont{Diehl and Shpot}(2000)}]{DiehlShpot00}
\bibinfo{author}{\bibfnamefont{H.~W.} \bibnamefont{Diehl}} \bibnamefont{and}
  \bibinfo{author}{\bibfnamefont{M.}~\bibnamefont{Shpot}},
  \bibinfo{journal}{Phys. Rev.~B} \textbf{\bibinfo{volume}{62}},
  \bibinfo{pages}{12338} (\bibinfo{year}{2000}).

\bibitem[{\citenamefont{Pleimling and Henkel}(2001)}]{PleimlingHenkel01}
\bibinfo{author}{\bibfnamefont{M.}~\bibnamefont{Pleimling}} \bibnamefont{and}
  \bibinfo{author}{\bibfnamefont{M.}~\bibnamefont{Henkel}},
  \bibinfo{journal}{Phys. Rev. Lett.} \textbf{\bibinfo{volume}{87}},
  \bibinfo{pages}{125702} (\bibinfo{year}{2001}).

\bibitem[{\citenamefont{Taylor and Györffy}(1993)}]{Taylor93}
\bibinfo{author}{\bibfnamefont{M.~B.} \bibnamefont{Taylor}} \bibnamefont{and}
  \bibinfo{author}{\bibfnamefont{B.~L.} \bibnamefont{Györffy}},
  \bibinfo{journal}{J.~Phys: Condens. Matter} \textbf{\bibinfo{volume}{5}},
  \bibinfo{pages}{4527} (\bibinfo{year}{1993}).

\bibitem[{\citenamefont{MacIsaac et~al.}(1995)\citenamefont{MacIsaac,
  Whitehead, Robinson, and De'Bell}}]{MacIsaac95}
\bibinfo{author}{\bibfnamefont{A.~B.} \bibnamefont{MacIsaac}},
  \bibinfo{author}{\bibfnamefont{J.~P.} \bibnamefont{Whitehead}},
  \bibinfo{author}{\bibfnamefont{M.~C.} \bibnamefont{Robinson}},
  \bibnamefont{and} \bibinfo{author}{\bibfnamefont{K.}~\bibnamefont{De'Bell}},
  \bibinfo{journal}{Phys. Rev.~B} \textbf{\bibinfo{volume}{51}},
  \bibinfo{pages}{16033} (\bibinfo{year}{1995}).

\bibitem[{\citenamefont{Hucht and Grüneberg}(2004)}]{HuchtGrueneberg04}
\bibinfo{author}{\bibfnamefont{A.}~\bibnamefont{Hucht}} \bibnamefont{and}
  \bibinfo{author}{\bibfnamefont{D.}~\bibnamefont{Grüneberg}}
  (\bibinfo{year}{2004}), \bibinfo{note}{in preparation}.

\bibitem[{\citenamefont{Fisher et~al.}(1972)\citenamefont{Fisher, k.~Ma, and
  Nickel}}]{FisherMaNickel72}
\bibinfo{author}{\bibfnamefont{M.~E.} \bibnamefont{Fisher}},
  \bibinfo{author}{\bibfnamefont{S.}~\bibnamefont{k.~Ma}}, \bibnamefont{and}
  \bibinfo{author}{\bibfnamefont{B.~G.} \bibnamefont{Nickel}},
  \bibinfo{journal}{Phys. Rev. Lett.} \textbf{\bibinfo{volume}{29}},
  \bibinfo{pages}{917} (\bibinfo{year}{1972}).

\bibitem[{\citenamefont{Fisher}(1983)}]{Fisher82}
\bibinfo{author}{\bibfnamefont{M.~E.} \bibnamefont{Fisher}}, in
  \emph{\bibinfo{booktitle}{Proceedings of the School on Critical Phenomena,
  Stellenbosch, South Africa, 1982}}, edited by
  \bibinfo{editor}{\bibfnamefont{F.~J.~W.} \bibnamefont{Hahne}}
  (\bibinfo{publisher}{Springer Verlag}, \bibinfo{address}{Berlin Heidelberg},
  \bibinfo{year}{1983}), vol. \bibinfo{volume}{186},
  chap.~\bibinfo{chapter}{1}.

\bibitem[{\citenamefont{Privman and Fisher}(1983)}]{PrivmanFisher83}
\bibinfo{author}{\bibfnamefont{V.}~\bibnamefont{Privman}} \bibnamefont{and}
  \bibinfo{author}{\bibfnamefont{M.~E.} \bibnamefont{Fisher}},
  \bibinfo{journal}{J.~Stat. Phys.} \textbf{\bibinfo{volume}{33}},
  \bibinfo{pages}{385} (\bibinfo{year}{1983}).

\bibitem[{\citenamefont{Lübeck and Heger}(2003)}]{LuebeckHeger2003a}
\bibinfo{author}{\bibfnamefont{S.}~\bibnamefont{Lübeck}} \bibnamefont{and}
  \bibinfo{author}{\bibfnamefont{P.~C.} \bibnamefont{Heger}},
  \bibinfo{journal}{Phys. Rev. Lett.} \textbf{\bibinfo{volume}{90}},
  \bibinfo{pages}{230601} (\bibinfo{year}{2003}).

\bibitem[{\citenamefont{Privman and Fisher}(1984)}]{PrivmanFisher84}
\bibinfo{author}{\bibfnamefont{V.}~\bibnamefont{Privman}} \bibnamefont{and}
  \bibinfo{author}{\bibfnamefont{M.~E.} \bibnamefont{Fisher}},
  \bibinfo{journal}{Phys. Rev.~B} \textbf{\bibinfo{volume}{30}},
  \bibinfo{pages}{322} (\bibinfo{year}{1984}).

\bibitem[{\citenamefont{Binder}(1981)}]{Binder81}
\bibinfo{author}{\bibfnamefont{K.}~\bibnamefont{Binder}},
  \bibinfo{journal}{Z.~Phys.~B} \textbf{\bibinfo{volume}{43}},
  \bibinfo{pages}{119} (\bibinfo{year}{1981}).

\bibitem[{\citenamefont{Binder et~al.}(1985)\citenamefont{Binder, Nauenberg,
  Privman, and Young}}]{BinNauPrivYoung85}
\bibinfo{author}{\bibfnamefont{K.}~\bibnamefont{Binder}},
  \bibinfo{author}{\bibfnamefont{M.}~\bibnamefont{Nauenberg}},
  \bibinfo{author}{\bibfnamefont{V.}~\bibnamefont{Privman}}, \bibnamefont{and}
  \bibinfo{author}{\bibfnamefont{A.~P.} \bibnamefont{Young}},
  \bibinfo{journal}{Phys. Rev.~B} \textbf{\bibinfo{volume}{31}},
  \bibinfo{pages}{1498} (\bibinfo{year}{1985}).

\bibitem[{\citenamefont{Brézin and Zinn-Justin}(1985)}]{BrezinZinnJustin85}
\bibinfo{author}{\bibfnamefont{E.}~\bibnamefont{Brézin}} \bibnamefont{and}
  \bibinfo{author}{\bibfnamefont{J.}~\bibnamefont{Zinn-Justin}},
  \bibinfo{journal}{Nucl. Phys.~B} \textbf{\bibinfo{volume}{257}},
  \bibinfo{pages}{687} (\bibinfo{year}{1985}).

\bibitem[{\citenamefont{Rudnick et~al.}(1985)\citenamefont{Rudnick, Guo, and
  Jasnow}}]{RudnickGuoJasnow85}
\bibinfo{author}{\bibfnamefont{J.}~\bibnamefont{Rudnick}},
  \bibinfo{author}{\bibfnamefont{H.}~\bibnamefont{Guo}}, \bibnamefont{and}
  \bibinfo{author}{\bibfnamefont{D.}~\bibnamefont{Jasnow}},
  \bibinfo{journal}{J.~Stat. Phys.} \textbf{\bibinfo{volume}{41}},
  \bibinfo{pages}{353} (\bibinfo{year}{1985}).

\bibitem[{\citenamefont{Aharony}(1976)}]{Aharony76}
\bibinfo{author}{\bibfnamefont{A.}~\bibnamefont{Aharony}}, in
  \emph{\bibinfo{booktitle}{Phase Transitions and Critical Phenomena}}, edited
  by \bibinfo{editor}{\bibfnamefont{C.}~\bibnamefont{Domb}} \bibnamefont{and}
  \bibinfo{editor}{\bibfnamefont{M.~S.} \bibnamefont{Green}}
  (\bibinfo{publisher}{Academic Press}, \bibinfo{address}{London},
  \bibinfo{year}{1976}), vol.~\bibinfo{volume}{6}, chap.~\bibinfo{chapter}{6}.

\bibitem[{\citenamefont{Glasser et~al.}(1987)\citenamefont{Glasser, Privman,
  and Schulman}}]{GlassPrivSchul87}
\bibinfo{author}{\bibfnamefont{M.~L.} \bibnamefont{Glasser}},
  \bibinfo{author}{\bibfnamefont{V.}~\bibnamefont{Privman}}, \bibnamefont{and}
  \bibinfo{author}{\bibfnamefont{L.~S.} \bibnamefont{Schulman}},
  \bibinfo{journal}{Phys. Rev.~B} \textbf{\bibinfo{volume}{35}},
  \bibinfo{pages}{1841} (\bibinfo{year}{1987}).

\bibitem[{\citenamefont{Wolff}(1989)}]{Wolff89}
\bibinfo{author}{\bibfnamefont{U.}~\bibnamefont{Wolff}},
  \bibinfo{journal}{Phys. Rev. Lett.} \textbf{\bibinfo{volume}{62}},
  \bibinfo{pages}{361} (\bibinfo{year}{1989}).

\bibitem[{\citenamefont{Ferrenberg and Swendsen}(1988)}]{Ferrenberg88}
\bibinfo{author}{\bibfnamefont{A.~M.} \bibnamefont{Ferrenberg}}
  \bibnamefont{and} \bibinfo{author}{\bibfnamefont{R.~H.}
  \bibnamefont{Swendsen}}, \bibinfo{journal}{Phys. Rev. Lett.}
  \textbf{\bibinfo{volume}{61}}, \bibinfo{pages}{2635} (\bibinfo{year}{1988}).

\bibitem[{\citenamefont{Aharony}(1983)}]{Aharony82}
\bibinfo{author}{\bibfnamefont{A.}~\bibnamefont{Aharony}}, in
  \emph{\bibinfo{booktitle}{Proceedings of the School on Critical Phenomena,
  Stellenbosch, South Africa, 1982}}, edited by
  \bibinfo{editor}{\bibfnamefont{F.~J.~W.} \bibnamefont{Hahne}}
  (\bibinfo{publisher}{Springer Verlag}, \bibinfo{address}{Berlin Heidelberg},
  \bibinfo{year}{1983}), vol. \bibinfo{volume}{186},
  chap.~\bibinfo{chapter}{3}.

\bibitem[{\citenamefont{Hucht et~al.}(1995)\citenamefont{Hucht, Moschel, and
  Usadel}}]{Hucht95}
\bibinfo{author}{\bibfnamefont{A.}~\bibnamefont{Hucht}},
  \bibinfo{author}{\bibfnamefont{A.}~\bibnamefont{Moschel}}, \bibnamefont{and}
  \bibinfo{author}{\bibfnamefont{K.~D.} \bibnamefont{Usadel}},
  \bibinfo{journal}{J.~Magn. Magn. Mater.} \textbf{\bibinfo{volume}{148}},
  \bibinfo{pages}{32} (\bibinfo{year}{1995}).

\bibitem[{\citenamefont{Hucht}(1998-2003)}]{fsscale}
\bibinfo{author}{\bibfnamefont{F.}~\bibnamefont{Hucht}},
  \emph{\bibinfo{title}{fsscale: A program for doing Finite-Size Scaling}}
  (\bibinfo{year}{1998-2003}),
  \urlprefix\url{http://www.thp.Uni-Duisburg.de/fsscale/}.

\bibitem[{\citenamefont{Luijten et~al.}(1999)\citenamefont{Luijten, Binder, and
  Blöte}}]{LuiBinBloe99}
\bibinfo{author}{\bibfnamefont{E.}~\bibnamefont{Luijten}},
  \bibinfo{author}{\bibfnamefont{K.}~\bibnamefont{Binder}}, \bibnamefont{and}
  \bibinfo{author}{\bibfnamefont{H.~W.~J.} \bibnamefont{Blöte}},
  \bibinfo{journal}{Eur. Phys. J. B} \textbf{\bibinfo{volume}{9}},
  \bibinfo{pages}{289} (\bibinfo{year}{1999}).

\bibitem[{\citenamefont{Kotz and Nadarajah}(2000)}]{KotzNadarajah00}
\bibinfo{author}{\bibfnamefont{S.}~\bibnamefont{Kotz}} \bibnamefont{and}
  \bibinfo{author}{\bibfnamefont{S.}~\bibnamefont{Nadarajah}},
  \emph{\bibinfo{title}{Extreme Value Distributions}}
  (\bibinfo{publisher}{Imperial College Press}, \bibinfo{address}{London},
  \bibinfo{year}{2000}).

\bibitem[{\citenamefont{Luijten and Blöte}(1996)}]{LuijtenBloete96}
\bibinfo{author}{\bibfnamefont{E.}~\bibnamefont{Luijten}} \bibnamefont{and}
  \bibinfo{author}{\bibfnamefont{H.~W.~J.} \bibnamefont{Blöte}},
  \bibinfo{journal}{Phys. Rev. Lett.} \textbf{\bibinfo{volume}{76}},
  \bibinfo{pages}{1557} (\bibinfo{year}{1996}).

\bibitem[{\citenamefont{Chen and Dohm}(1998)}]{ChenDohm98a}
\bibinfo{author}{\bibfnamefont{X.~S.} \bibnamefont{Chen}} \bibnamefont{and}
  \bibinfo{author}{\bibfnamefont{V.}~\bibnamefont{Dohm}},
  \bibinfo{journal}{Int. J.~Mod. Phys.~C} \textbf{\bibinfo{volume}{9}},
  \bibinfo{pages}{1007} (\bibinfo{year}{1998}).

\bibitem[{\citenamefont{Wolfram}(1999)}]{MMA4}
\bibinfo{author}{\bibfnamefont{S.}~\bibnamefont{Wolfram}},
  \emph{\bibinfo{title}{The Mathematica Book}} (\bibinfo{publisher}{Wolfram
  Media, Cambridge University Press}, \bibinfo{year}{1999}),
  \bibinfo{edition}{4th} ed.

\end{thebibliography}

\end{document}